\documentclass[twocolumn]{aastex631}
\usepackage{amsmath}

\newcommand{\ha}{H$\alpha$}

\newcommand{\photoz}{\(z_{photo}\)}
\newcommand{\mzph}{\(\tilde{z}_{photo}\)}

\begin{document}

\title{Mapping the Galaxy Color-Star Formation Rate Relation with Manifold Learning and Infrared Image Stacking }

\author[0000-0001-8792-3091]{Yu-Heng Lin}
\affiliation{Caltech/IPAC, 1200 E. California Blvd. Pasadena, CA 91125, USA}
\email{ianlin@ipac.caltech.edu}

\author[0000-0001-5382-6138]{Daniel Masters}
\affiliation{Caltech/IPAC, 1200 E. California Blvd. Pasadena, CA 91125, USA}

\author[0000-0002-9382-9832]{Andreas L. Faisst}
\affiliation{Caltech/IPAC, 1200 E. California Blvd. Pasadena, CA 91125, USA}

\author[0000-0002-7064-5424]{Harry Teplitz}
\affiliation{Caltech/IPAC, 1200 E. California Blvd. Pasadena, CA 91125, USA}

\author[0000-0002-7303-4397]{Olivier Ilbert}
\affiliation{Aix Marseille Univ., CNRS, CNES, LAM, Marseille, France}

\author[0000-0002-3915-2015]{ Matthieu Bethermin}
\affiliation{Université de Strasbourg, CNRS, Observatoire astronomique de Strasbourg, UMR 7550, F-67000 Strasbourg, France}
\affiliation{Aix Marseille Univ., CNRS, CNES, LAM, Marseille, France}

\author[0000-0003-2226-5395]{Shoubaneh Hemmati}
\affiliation{Caltech/IPAC, 1200 E. California Blvd. Pasadena, CA 91125, USA}

\author[0000-0001-7166-6035]{Vihang Mehta}
\affiliation{Caltech/IPAC, 1200 E. California Blvd. Pasadena, CA 91125, USA}

\author[0000-0002-4485-8549]{Jason D. Rhodes}
\affiliation{Jet Propulsion Laboratory, California Institute of Technology, 4800, Oak
Grove Drive, Pasadena, CA, 91011, USA}

\author[0000-0002-6313-6808]{Gregory L. Walth}
\affiliation{Caltech/IPAC, 1200 E. California Blvd. Pasadena, CA 91125, USA}

\begin{abstract}

Modern surveys present us with billions of faint galaxies for which we only have broadband images in $\sim$6-8 optical-to-near-infrared (NIR) filters.  Galaxy star formation rates (SFRs) are difficult to estimate accurately without spectroscopic diagnostics or far-infrared (FIR) photometry, both of which are prohibitively expensive to obtain for large numbers of faint, high-redshift galaxies. 
Here we present the empirical relation between SFR and broadband optical-to-NIR colors learned from Spitzer MIPS and Herschel PACS/SPIRE imaging using an innovative stacking analysis that bins galaxies with similar optical-to-NIR spectral energy distributions using a Self-Organizing Map (SOM). Stacking based on optical-to-NIR colors ensures that our FIR stacks are built from galaxies with similar intrinsic physical properties as opposed to stacking simply by stellar mass.  We train a 40$\times$40 SOM using 230,638 galaxies selected from the  COSMOS field, and stack the mid-to-far infrared images from 24 micron to 500 micron.  We are able to measure the median FIR luminosities from half of the SOM cells to calibrate the star formation rate. In addition to investigating the common structures of optical-to-NIR properties and FIR detections labeled on the SOM,  we provide calibrated star formation rates for nearly half of the galaxies in the COSMOS fields down to $i-$band magnitude $\leq 25.5$, and present the evolution of the galaxy main sequence for low-mass galaxies to redshift $z\sim2.5$.

\end{abstract}


%%%%%%%%%%%%%%%%%%%%%%%%%%%%%%%%%%%%%%%%%%%%%%%%%%%%%%%%%%%%%%%%%%
%%%%%%%%%%%%%%%%%%%%%%%%%%%%%%%%%%%%%%%%%%%%%%%%%%%%%%%%%%%%%%%%%%
\section{Introduction} \label{sec:intro}
%%%%%%%%%%%%%%%%%%%%%%%%%%%%%%%%%%%%%%%%%%%%%%%%%%%%%%%%%%%%%%%%%%
%%%%%%%%%%%%%%%%%%%%%%%%%%%%%%%%%%%%%%%%%%%%%%%%%%%%%%%%%%%%%%%%%%
The star formation rate (SFR), which sets the stellar mass buildup of galaxies, is among the most important quantities giving insight into galaxy evolution across cosmic time. The SFR depends on the nature of star-forming regions, gas inflows, mergers, and stellar and AGN feedback. However, measuring SFR is difficult. Spectroscopy can be used to derive the SFR from, e.g., the \ha\ recombination line, but it is time-consuming for faint galaxies, and the results depend on being able to adequately account for dust extinction \citep{Moustakas_2006, Kennicutt_2012}.
In addition, photometric colors can be used to estimate the \ha~luminosity, which in turn can be converted to SFRs \citep{shim_2011,faisst_2016,smit_2016}, however, this method is again suffering from uncertain nebular dust attenuation correction.
The rest-UV emission of galaxies can also be used to estimate SFR due to its correlation with the presence of young, UV-bright stars generated in recent star formation episodes \citep{K98}, but the emerging UV emission as well is highly sensitive to dust attenuation, making these estimates uncertain without good constraints on the amount of radiation that has been reprocessed into the far-infrared (FIR) by dust. 

SFR estimates based on both UV and FIR emission are more robust \citep{Kennicutt_2012} and, with missions like Herschel \citep{Pilbratt_2010}, are accessible for many galaxies. However, due to sensitivity limits, the galaxies individually detected by Herschel are strongly biased to high SFR (and also stellar masses) at all redshifts \citep{Pannella_2015}.   
Stacking images of galaxies in the same range of redshift and stellar mass is commonly used to estimate the FIR luminosity \citep{Schreiber_2015, Bethermin_2020, Leslie_2020, Leroy_2024}. This method is powerful for inferring the average SFR of a complete galaxy population as well as the average cosmic volume SFR density.  However, binning by stellar mass indifferent the galaxy photometric colors, which encode many physical properties including the SFR, therefore this stacking method is less ideal for individual galaxy SFR estimation.

An alternate approach is to derive SFRs from SED fitting of stellar population synthesis models to observed ultra-violet (UV), optical, near-infrared (NIR) colors, but SFR estimates from template fitting are generally considered imprecise and subject to large systematic errors arising from uncertainties in reddening, the initial mass function (IMF), or the star formation history prescription \citep{K98,Papovich_2001,Conroy_2013}. 
While major strides have been made in improving galaxy spectral templates, from both stellar population synthesis models \citep{BC_2003,Eldridge_2017} and empirical observations \citep{Brown_2014}, it is still the case that there are significant uncertainties preventing a confident determination of SFR from template fitting. 
This means that the varieties of the models and templates do not account for all the observational parameters.

In \citet{Masters_2015}, a technique was introduced to empirically constrain the relation between galaxy colors and redshift, motivated by the need to obtain unbiased photometric redshift (\photoz) estimates for weak lensing cosmology \citep{Ma_2006,Abdalla_2008,Laureijs_2011,Newman_2015}. The first part of this work was to quantify the actual distribution of galaxies in the high-dimensional color space relevant to \photoz\ estimation for Euclid and Roman. An observation made in \citet{Masters_2015} is that the galaxy color manifold is finite and measurable. In other words, galaxy colors are not random, and the ``same'' galaxy-class shows up over and over in a deep survey.

The self-organizing map \citep[SOM;][]{kohonen_1982} was used in \citet{Masters_2015} to measure the galaxy color manifold. The SOM is an unsupervised machine learning method designed to learn the distribution of high dimensional data and represent it with a finite number of topologically ordered basis vectors.
Manifold learning with the SOM helps to overcome the “curse of dimensionality” – that is, the exponential growth of the volume of the data space with the number of dimensions. Color cuts have been used successfully in astronomy for decades; manifold learning with the SOM expands the power of color cuts to higher color dimensions than the 2-3 to which they are typically applied.  
In the SOM framework, galaxies with similar input features are distributed to their ``best-matching unit" or ``best-matching cell".  Galaxies assigned to the same cell are treated as comparable sources after proper normalization. The SOM is then calibrate by labeling the cells with the median or normalized properties of the galaxies in the cells.   The calibrated SOM can be used for prediction by projecting new data onto the SOM \citep{Davidzon_2022}. 
The technique of SOM trained on the galaxy colors has been successfully applied to calibrate redshift \citep{Masters_2015,Hemmati2019a,Wright_2020,McCullough2024}, as well as the physical properties of the galaxies and AGNs \citep{faisst_2019,Hemmati2019,Davidzon_2019, Davidzon_2022, Sanjaripour_2024, LaTorre_2024,Jafariyazani_2025, Abedini_2025}. 
Here, we will extend the SOM methodology from labeling the cell with known properties to enhancing unseen signals via stacking FIR images.

In this work, we study the empirical relation between the galaxy FIR star-formation rate and broadband optical/NIR colors with the aid of manifold learning and stacking analysis. As opposed to binning galaxies by mass during the stacking process, we will bin the galaxies by their spectral energy distribution (SED) to discriminate the encoded physical properties. 
We cluster the galaxies in the field of the Cosmic Evolution Survey \citep[COSMOS;][]{Koekemoer_2007, Scoville_2007} with their optical/NIR broadband SED \citep{Weaver_2022} using the SOM, where galaxies assigned to the cell of the SOM have similar colors and redshifts.
In each unit, we stack the Spitzer and Herschel imaging of these galaxies. We then measure the FIR luminosity and the FIR star-formation rate from the stacked FIR imaging. 

This paper is structured as follows.  In Section~\ref{sec:data} we describe the object selection in photometric catalogs and FIR images in the COSMOS field.  In Section~\ref{sec:method}, we describe the methods we use to construct the color-SED self-organizing map, stack the FIR images, and measure the FIR luminosities across the color manifold.  In Section~\ref{sec:result}, we present and discuss the FIR photometry and SFR we inferred from the SOM stacking analysis. The summary is in Section~\ref{sec:summary}.
Throughout this work, we assume a $\Lambda$CDM cosmology with $H_0=$70km s$^{-1}$ Mpc$^{-1}$, $\Omega_\Lambda=$0.7, and $\Omega_m=$0.3. All magnitudes are given in the AB system.

%%%%%%%%%%%%%%%%%%%%%%%%%%%%%%%%%%%%%%%%%%%%%%%%%%%%%%%%%%%%%%%%%%
%%%%%%%%%%%%%%%%%%%%%%%%%%%%%%%%%%%%%%%%%%%%%%%%%%%%%%%%%%%%%%%%%%
\section{Data} \label{sec:data}
%%%%%%%%%%%%%%%%%%%%%%%%%%%%%%%%%%%%%%%%%%%%%%%%%%%%%%%%%%%%%%%%%%
%%%%%%%%%%%%%%%%%%%%%%%%%%%%%%%%%%%%%%%%%%%%%%%%%%%%%%%%%%%%%%%%%%

\subsection{Optical and Near-Infrared Photometry}
We train our SOM from the galaxy SED's adjacent colors $u-g$, $g-r$, $r-i$, $i-z$, $z-Y$, $Y-J$, $J-H$ for galaxies in the Cosmic Evolution Survey \citep[COSMOS;][]{Koekemoer_2007, Scoville_2007} field. These bands were chosen as being representative of the large-scale data that is or will be available from LSST \citep{LSST}, Euclid \citep{Euclid}, and Roman \citep{Spergel_2015,Akeson_2019}.
The $u$ band was measured by the Canada-France-Hawaii telescope (CFHT) program CLAUDS \citep{Sawicki_2019}. 
The $g, r, i, z$ bands were measured with the Subaru Hyper-Suprime-Cam \citep[HSC]{Miyazaki_2018} from the second public data release of the HSC Subaru Strategic Program \citep[HSC-SSP, ][]{Aihara_2019}. 
The $Y, J, H$ bands were measured from the UltraVISTA program using VIRCAM on the VISTA telescope \citep{McCracken_2012}. 
We adopt the 2$^{"}$ aperture photometry (\texttt{MAG\_APER2}) in the COSMOS2020 CLASSIC catalog \citep{Weaver_2022} for the galaxy magnitudes and colors.  The physical properties in the catalog, such as photometric redshift, stellar mass, and optical-to-NIR SED star formation rate, are calculated from the fitting code \texttt{LePhare} \citep{Arnouts_1999, Ilbert_2006} as described in detail in \citet{Weaver_2022}.  The \photoz in the COSMOS2020 catalog have sub-percent dispersion for bright sources ($i <$21), and $<5\%$ for fainter objects \citep{Weaver_2022, Khostovan_2025}.   

To assemble our training data, we applied the following criteria: 
\begin{align*}
\texttt{HSC\_i\_MAG\_APER2} & < 25.5, \\
\texttt{UVISTA\_J\_MAG\_APER2} & < 24.5, \\
\texttt{lp\_type} & = 0.
%abs(\texttt{A\_MAG\_APER2} - \texttt{B\_MAG\_APER2}) & > 0 , \\
%abs(\texttt{A\_MAGERR\_APER2} - \texttt{B\_MAGERR\_APER2}) & > 0.
\end{align*}
The $\texttt{MAG\_APER2}$ label represent the 2$^{"}$ aperture photometry magnitude.  
%The $\texttt{A\_MAG\_APER2}$ and $\texttt{A\_MAGERR\_APER2}$ represent the 2$^{"}$ aperture photometry magnitude and the uncertainty of the broadband filter $\texttt{A}$. The filter $\texttt{B}$ is the adjacent broadband of filter $\texttt{A}$.
The $\texttt{lp\_type} \leq 0 $ are objects classified as galaxies (excluding sources of stars, X-ray sources, and failure in fit). 
We also request the photometry magnitudes and uncertainties reported as numerical values to generate the adjacent colors and uncertainties. 
As the result, we have 230,637 galaxies for our training sample. 
Galaxies at redshift $z\gtrsim2.5$ may have undetected $u-$bands due to the rest-frame UV drop off.  We do not mask the magnitude values for the null detection.

\subsection{Far-Infrared Images and Photometry}
The FIR images of the stacking analysis come from the Multiband Imaging Photometer for Spitzer \citep[MIPS;][]{Rieke_2004} in the S-COSMOS program \citep{Sanders_2007}, the Herschel Photodetector Array Camera and Spectrometer \citep[PACS;][]{Poglitsch_2010} in the PACS Evolutionary Probe \citep[PEP;][]{Lutz_2011}, and the Herschel  Spectral and Photometric Imaging Receiver \citep[SPIRE;][]{Griffin_2010} in the Herschel Multi-tiered Extragalactic Survey  \citep[HerMES;][]{Oliver_2012}.
The Spitzer MIPS imaging in S-COSMOS is comprised of three broad spectral bands centered at 24 and 70 $\mu$m.
The Herschel PACS imagings in PEP are centered at 100 $\mu$m and 160 $\mu$m.
The Herschel SPIRE imaging wavelengths in HerMES are centered at 250, 350, and 500 $\mu$m.
We apply the point spread function (PSF) fitting for the photometry.  In particular, we follow the photometry method in \citet{Schreiber_2015}, where the 0.9$\times$ the full-width-half-maximum (FWHM) of the PSF are fitted to the source images ( see Section~\ref{sec:stacking} for more detail). 
We listed the image characteristics in Table~\ref{tab:data_tab}.

\begin{table}[ht]
\centering
\begin{tabular}{cccc}
\hline\hline
\vspace{1 mm}
 \shortstack{Band \\ ($\mu$m)} & \shortstack{Image pixel scale \\ ($''$/pixel)} & \shortstack{PSF FWHM \\ ($''$)} & \shortstack{$3\sigma$ depth \\ (mJy)} \vspace{2 mm}
 \\
24  & 1.2  & 6.4   & 0.071 \\
70  & 4.0   & 16.0   & 7.5 \\
100 & 1.2   & 6.8   & 5.0 \\
160 & 2.4   & 11.4  & 10.2 \\
250 & 6.0   & 18.2  & 8.1 \\
350 & 8.33  & 24.9  & 10.7 \\
500 & 12.0  & 36.3  & 15.4 \\
\hline
\end{tabular}
\caption{ Image characteristics of Spitzer/MIPS 24 and 70 $\mu$m, Herschel PACS 100 and 160 $\mu$m, and SPIRE  250, 350, and 500 $\mu$m images in the COSMOS.}
\label{tab:data_tab}
\end{table}

%%%%%%%%%%%%%%%%%%%%%%%%%%%%%%%%%%%%%%%%%%%%%%%%%%%%%%%%%%%%%%%%%%
%%%%%%%%%%%%%%%%%%%%%%%%%%%%%%%%%%%%%%%%%%%%%%%%%%%%%%%%%%%%%%%%%%
\section{Method} \label{sec:method}
%%%%%%%%%%%%%%%%%%%%%%%%%%%%%%%%%%%%%%%%%%%%%%%%%%%%%%%%%%%%%%%%%%
%%%%%%%%%%%%%%%%%%%%%%%%%%%%%%%%%%%%%%%%%%%%%%%%%%%%%%%%%%%%%%%%%%

\subsection{Constructing the Self Organizing Map}

The SOM is an unsupervised machine learning method to project high-dimensional data onto a lower-dimensional grid \citep{kohonen_1982}. This means that the SOM is agnostic to its input and therefore does not need a training or test set, unlike the supervised machine learning methods.  Similar objects are grouped together on the SOM, and structures that exist in the high dimensional data space are preserved in the lower-dimensional latent space.

In the SOM training process, each cell is assigned an initialized weight vector.  Our input vectors $\mathbf{x}$ are the 7 features in the color SED: $\mathbf{x}=(u-g,\,g-r,\,r-i,\,i-z,\,z-Y,\,Y-J,\,J-H)$, which is effectively a normalized SED.  We find the best-matching cell by searching for the minimum distance $min( d_c )$ between the input vector $\mathbf{x}$ and the cells' weight vector $\mathbf{w}_c$, where the distance is defined as the {\it Euclidean} distance divided by the uncertainties $\Delta \mathbf{x}$ in magnitude: 
\begin{equation}\label{eq:distance}
d_c^2(\mathbf{x}, \mathbf{w}_c) = \sum^{m=7}_{i=1}( \frac{x_j - w_{c,i}}{\Delta x_i} )^2.
\end{equation}
We then update the weight vectors of the best-matching-cell and neighbors of the next matching cell in each iteration.
After 200,000 iterations, the galaxies with similar SED will be assigned to the same cell on the SOM, and cells with similar weight vectors (effectively the SED representing the cell) are placed nearby on the SOM.
The training process and structure of the SOM in this work are detailed in \citet{Masters_2015}. 

Throughout the studies of clustering galaxies using SOM, the input vector can be magnitudes, colors, or both magnitudes and colors \citep{Wright_2020}.  In general, the magnitudes are sensitive to the scale of the galaxy, such as mass and luminosity, and the colors are sensitive to the stellar components and dust properties. 
If the input vector is merely the magnitudes, the clustering will prioritize scales over the color-related properties.
If the input vector includes both the magnitude and the colors, then both the scales and color-related properties are considered. Having more features in the input vector can improve the clustering performance when using a larger size of SOM, which has more cells to handle the finer classification. However, in this work, we have to balance between having enough galaxies per cell for the FIR stacking analysis while maintaining low dispersion in galaxy properties per cell; therefore, we choose the color SED as our input vector to prioritize the discrimination of color-related properties.  
After testing the SOM sizes, we group the broad-band color SEDs of 230,637 galaxies on the SOM with dimensions of 40$\times$40 cells.

\subsubsection{Galaxy properties across the SOM}

Since the SOM is an unsupervised dimensionality reduction method, it can be better understood by labeling each cell with the assigned galaxies' statistics, such as the median and dispersion of the redshifts, magnitude, etc. 
Since the galaxies with similar SED are assigned to the same cell, we expect the physical properties relevant to the input data, such as photometric redshift, to have a small dispersion within the cell. Likewise, since each cell is surrounded by cells with similar weight vectors, median values of the galaxy properties labeled to the cells show structures on the SOM, and the same structures may appear in maps colored by different properties.

\begin{figure}
\includegraphics[width=0.98\linewidth]{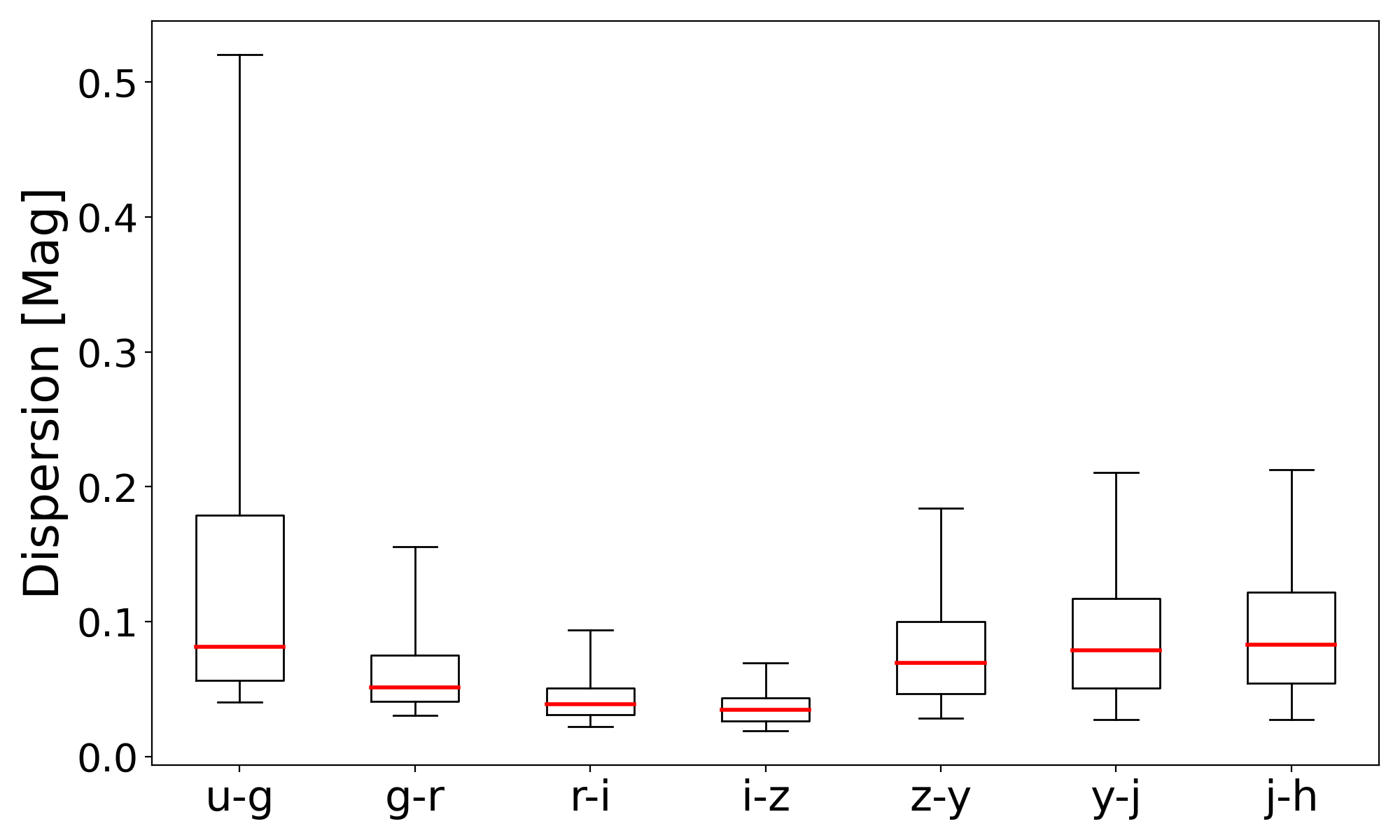}
\caption{  The median absolute dispersion of the adjacent photometric colors.  The red lines mark the median, and the upper and lower limits of the bars and the boxes are (95, 5) percentile and (75, 25) percentile, respectively. 
\label{fig:mag_dispersion}}
\end{figure}

In Figure~\ref{fig:mag_dispersion}, we show the median absolute dispersions of input features labeled to the SOM cells in the box-plots. The red lines in the boxes are the median values, and the upper and lower limits of the bars and the boxes are (95, 5) and (75, 25) percentiles of their dispersion, respectively.  The dispersions of the colors in the SOM cells are typically 0.02-0.07 magnitude in the optical and $0.03-0.15$ magnitude in the NIR.  The larger dispersions of the $u-g$ color are caused by the unconstrained magnitudes of non-detections in $u-$band.  Since we do not apply any value-masking for non-detections, the undetected $u-$band will result to a large $u-g$ value, making $u-g$ a distinct feature from data with detected $u-$band. But the uncertainty of $u-g$ will also be large, which reduces the significance of the $u-g$ feature while comparing between the galaxies with undetected $u-$band.  

\begin{figure}
\includegraphics[width=0.98\linewidth]{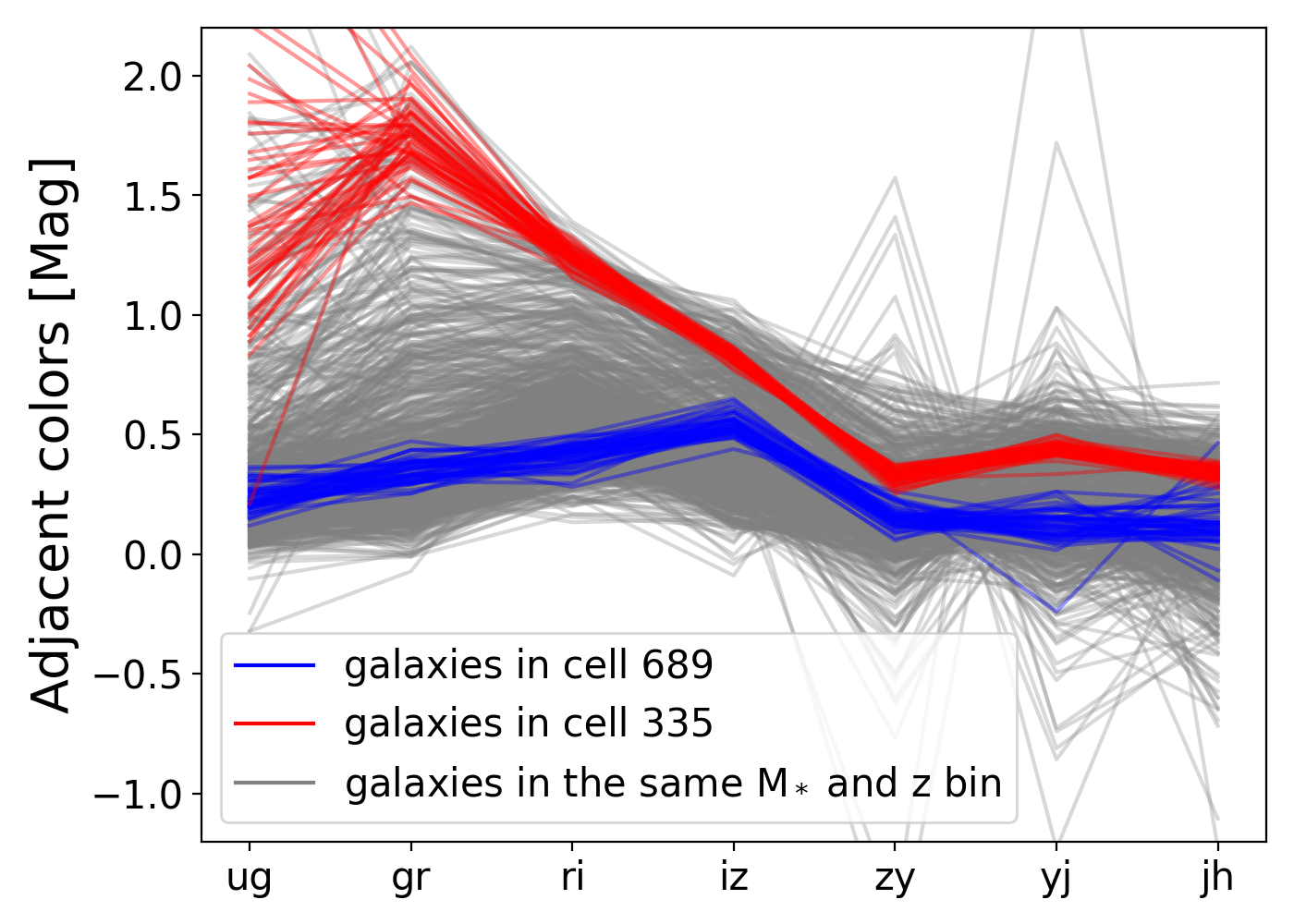}
\caption{  The color SEDs of galaxies within the bin of $10.5<\log(M_*/{\rm M_\odot})<11.0$ and $0.9<z<1.2$.   The gray lines are the SEDs of 800 galaxies randomly selected from this mass-redshift bin.  The red and blue lines are the SEDs of 100 galaxies from the same mass-redshift bin, but assigned to different cells on the SOM.   
\label{fig:SED_mz_vs_SOM}}
\end{figure}

In Figure~\ref{fig:SED_mz_vs_SOM}, we highlight our motivation for clustering and stacking (Section~\ref{sec:stacking}) the galaxies by color SED on the SOM in contrast to over the stellar mass and redshift.  Binning galaxies by stellar mass and redshift bin will group together galaxies with diverse characteristics, while our method will be sensitive to the differences between these galaxies and only accumulate the similar ones. It will therefore lead to a ``cleaner'' way of stacking compared to other methods using just redshift and stellar mass binning.

\begin{figure}
\includegraphics[width=0.98\linewidth]{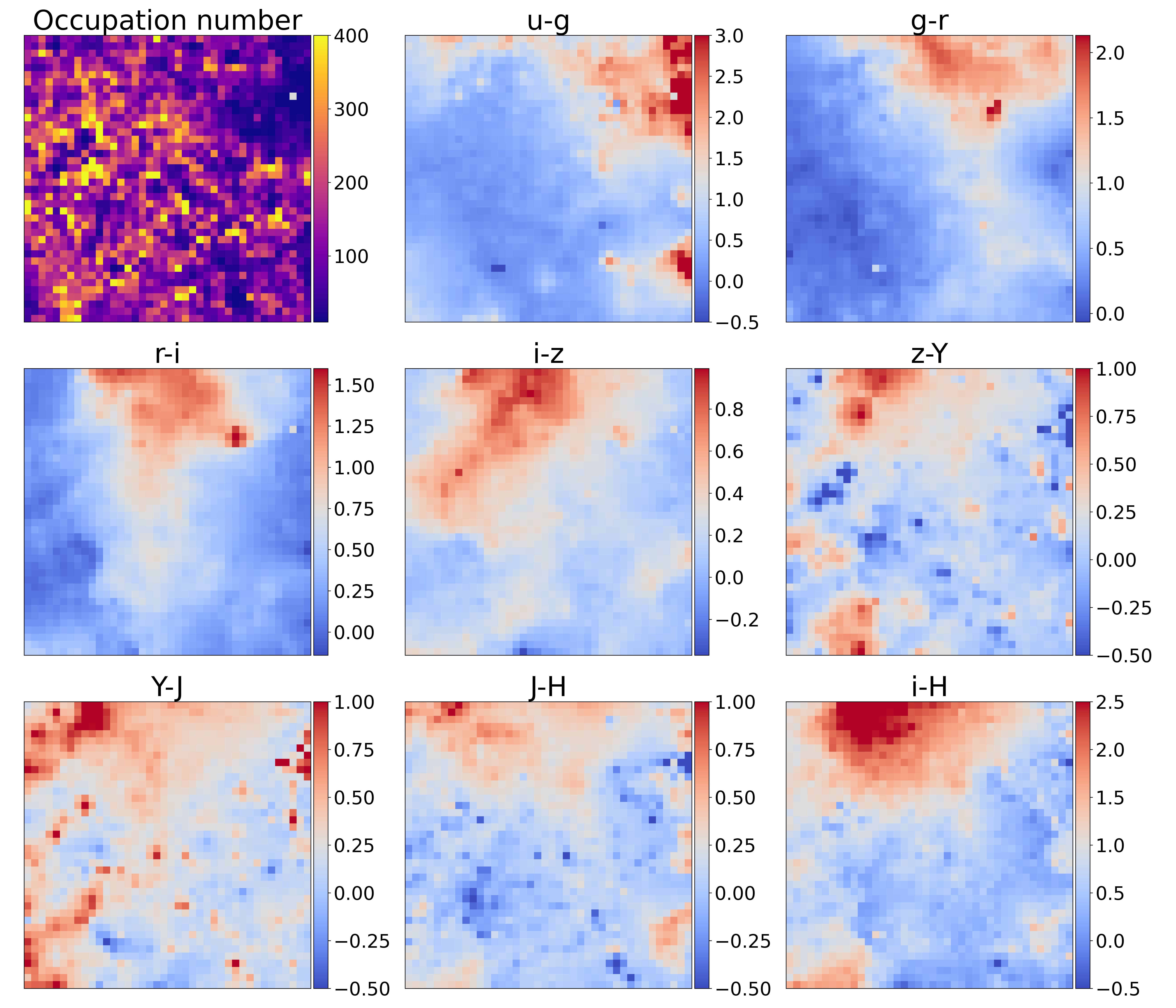}
\caption{  The 40$\times$40 cell SOM of COSMOS galaxies labeled with the occupation numbers per cell, the adjacent optical-to-NIR photometric colors used to train the map, and the $i-H$ color.  
\label{fig:labeled_SOM1}}
\end{figure}

In Figure~\ref{fig:labeled_SOM1}, we first show the SOM labeled with the occupation numbers per cell, the input adjacent colors, and $i-H$ color as NIR red color. The map labeled with the occupation numbers shows no specific structure similar to maps labeled with the FIR stacking results (shown later in Figure~\ref{fig:SOM_mosaic}) , which means that the detection of the FIR stacked images are not driven only by the numbers of stacks.    

\begin{figure*}
\includegraphics[width=0.98\linewidth]{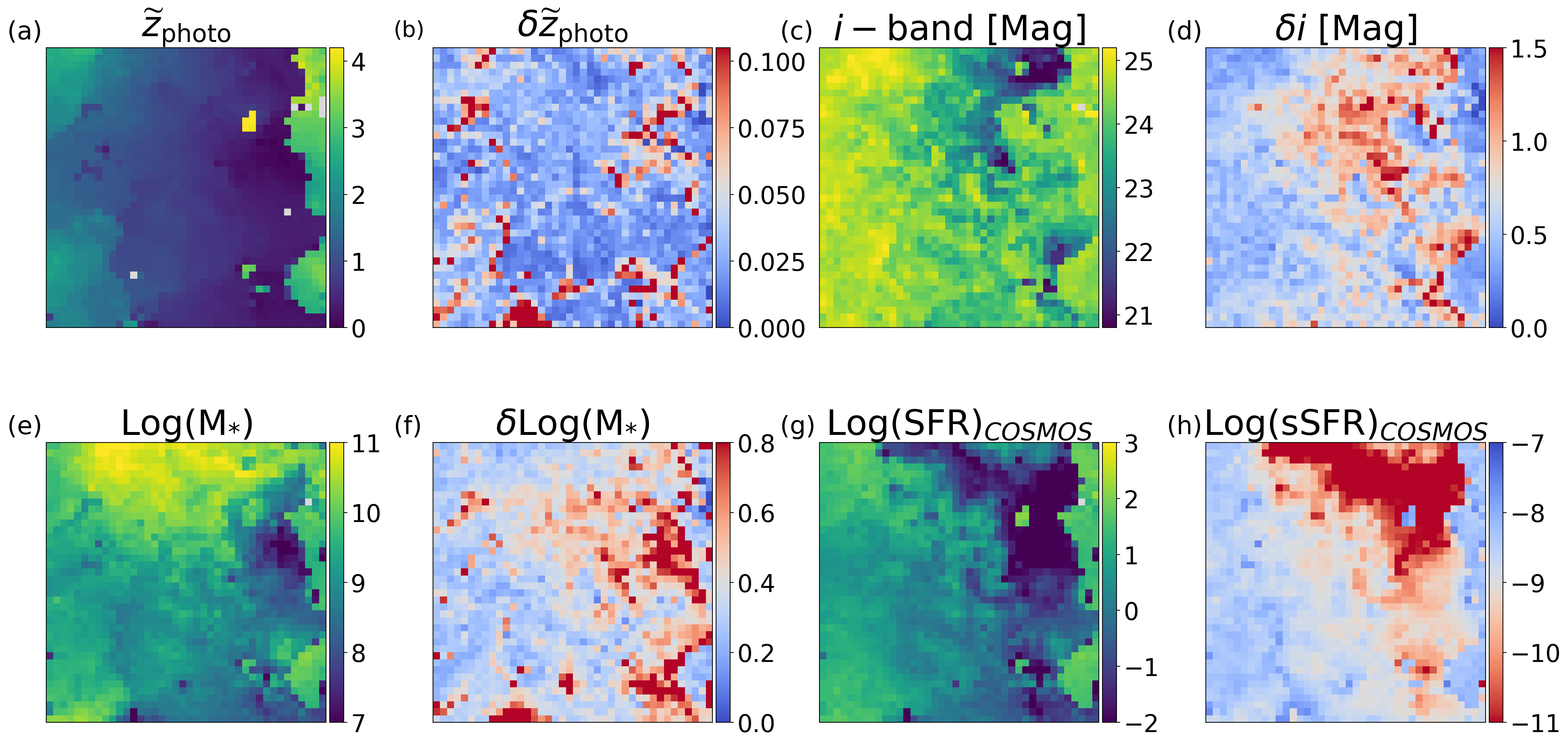}
\caption{   The 40$\times$40 SOM of COSMOS galaxies labeled with (a) the median of photometric redshifts, (b) the redshift dispersions, (c) the assigned galaxies' median of $i-$band magnitudes, (d) the dispersions of the $i-$band magnitudes, (e) the median of Log($M_*$), (f) the Log($M_*$) dispersions, (g) the median of Log(SFR),(h) the median of Log(sSFR)of the galaxies in each cell.  The $z_{photo}$, $M_*$, SFR and sSFR are adopted from the COSMOS2020 classic catalog \citep{Weaver_2022}.
\label{fig:labeled_SOM2}}
\end{figure*}

In Figure~\ref{fig:labeled_SOM2}, we first show the SOM labeled with the \photoz\ median, denoted as $\tilde{z}_{photo}$. These \photoz, adopted from the COSMOS2020 catalog, have sub-percent dispersion for bright sources ($i <$21), and $<5\%$ for fainter objects \citep{Weaver_2022, Khostovan_2025}.   
The redshifts of the cells range from $\tilde{z}_{photo}$ = 0.014 to $\tilde{z}_{photo}$ = 4.4044.  The \photoz\ median on the cells varies smoothly throughout most parts of the SOM, while some sharp edges are formed with redshift differences of $\Delta z\sim3$. 
This is caused by the similar photometric colors of the Lyman break in high redshift galaxies and the 4000\AA~break in low redshift galaxies when the redshift difference is around 3.

We define the redshift dispersion as $\delta z \equiv NMAD(z)/(1+\tilde{z}_{photo})$, where $NMAD(z)$ is the normalized median absolute deviation of the redshifts.
In our SOM, 97$\%$ of the cells have  $\delta z < 0.15$, and only 15 cells (0.9$\%$ of the whole SOM) have $\delta z > 0.3$. The cells with large redshift dispersion are either located where the redshift transitions sharply, as shown in Figure~\ref{fig:labeled_SOM2} panel (b), or with very low occupation numbers (see Figure~\ref{fig:flag_scatter1}). 

The sharp transitions on the redshift-labeled SOM also appear in several color-labeled SOMs in Figure~\ref{fig:labeled_SOM1}, such as the $u-g$ and $r-i$ maps, since these colors cover the galaxy SED features that are informative for deciding the redshifts, such as the Lyman break and the 4000 \AA\ break.  

We also show the SOM labeled with the observed $i-$band magnitude, Log(M$_*$), the logarithm dispersion $\delta$Log(M$_*$) in Figure~\ref{fig:labeled_SOM2}, together with the logarithm of SFR and specific star formation rate (sSFR) adopted in COSMOS2020. We find that star-forming galaxies with sSFR $>=10^{-9}$ yr$^{-1}$ are separated from the quiescent galaxies (sSFR $<10^{-9}$ yr$^{-1}$). 
The $i-band$ magnitude dispersions $\delta i$ are small when the median $i-band$ magnitude are faint, and gradually increase on the region where the median $i-band$ magnitudes are brighter.   
The small dispersions are due to the selection bias introduced by the magnitude cut in Section~\ref{sec:data}.  The cells that gather low redshift galaxies (hence brighter $i-$band) will include a wider range of galaxy brightness, resulting to and a larger dispersion.  Whereas in the cells that gather higher redshift galaxies (fainter $i-$band), the range of galaxy brightness are chopped by our magnitude cut, resulting to a smaller dispersion.  The stellar mass dispersions are affected the same way as $i-band$ magnitudes dispersion.  
The stellar mass dispersion are also higher at the region of sharp redshift transition. That is because the cells at the sharp redshift transition contain galaxies with different $z_{photo}$, whose stellar mass are derived from similar observed SED but distinct rest-frame SED.

\subsection{Stacking the Far-Infrared Images}\label{sec:stacking}

To perform stacking of FIR images for galaxies in each cell, we first make cutouts of the FIR images of the sources centered at their optical coordinates. The cutouts' size is chosen to be 100$\times$100 pixels, which is more than sufficient to capture both the source signal and background level in all IR bands used for this analysis (see the pixel scales in Table~\ref{tab:data_tab}). 
We then divide each image by its scaling factor $S$, where: 
\begin{equation}\label{eq:scaling_factor}
\log(S) = {\frac{( \text{med}( w_{c,i} - x_i ) )}{-2.5}},
\end{equation}
where $\text{med}( w_{c,i} - x_i )$ is the median of the magnitude differences between the best-matching-cell weight vector $\mathbf{w}_c$ and the source color vector $\mathbf{x}$. 
This normalizing process scales the FIR images based on the optical-to-NIR magnitudes difference between the cell representation and the source.
This process aims to align the contribution of bright and faint objects during the stacking (because we note that the SOM was only trained on colors and not on magnitudes themselves).  
When predicting the FIR photometry of a single galaxy, we can multiply its scaling factor by the average photometry of its best-matching cell. 
We find the scaling factors for the sources roughly follow a log-normal distribution, with $\log(S) = 0.0 \pm 0.30$.

We rotate the image by a random angle to remove potential oriented bias in the respective images.  
We stack these scaled images of galaxies in each cell by taking the median value pixel-wise, then chop the stacked image to the size of (50, 50) to remove the artifact edge caused by stacking the rotated images.  
For the background subtraction, we follow the PSF-fitting method in \citet{Schreiber_2015}, where we calculate the background of the stacked image while measuring the flux density using the central 0.9$\times$FWHM of the PSF.   
We have tested that removing the background before or after stacking does not significantly affect the flux density measurement. 
The Herschel fluxes are typically overestimated because of the clustering/blending effects from nearby galaxies. The bias of flux is wavelength dependent.  While PACS 100 and 160$\mu$m have modest blending, the bias in SPIRE images are higher due to larger beam size \citep{Bethermin_2017}. 
Here we correct the wavelength-dependent contamination correction using Table B.2 in \citet{Schreiber_2015}, as we are following their photometry measurement methods for our stacked images.    
In Section~\ref{sec:simulation}, we generate mock sources to simulate the stacking process and study the effects of scaling the FIR images with equation~\ref{eq:scaling_factor}. We also discuss the difference between the mean- and median-stacking methods in our work in Appendix~\ref{sec:bias_in_simulation}. 

\begin{figure}
\includegraphics[width=0.99\linewidth]{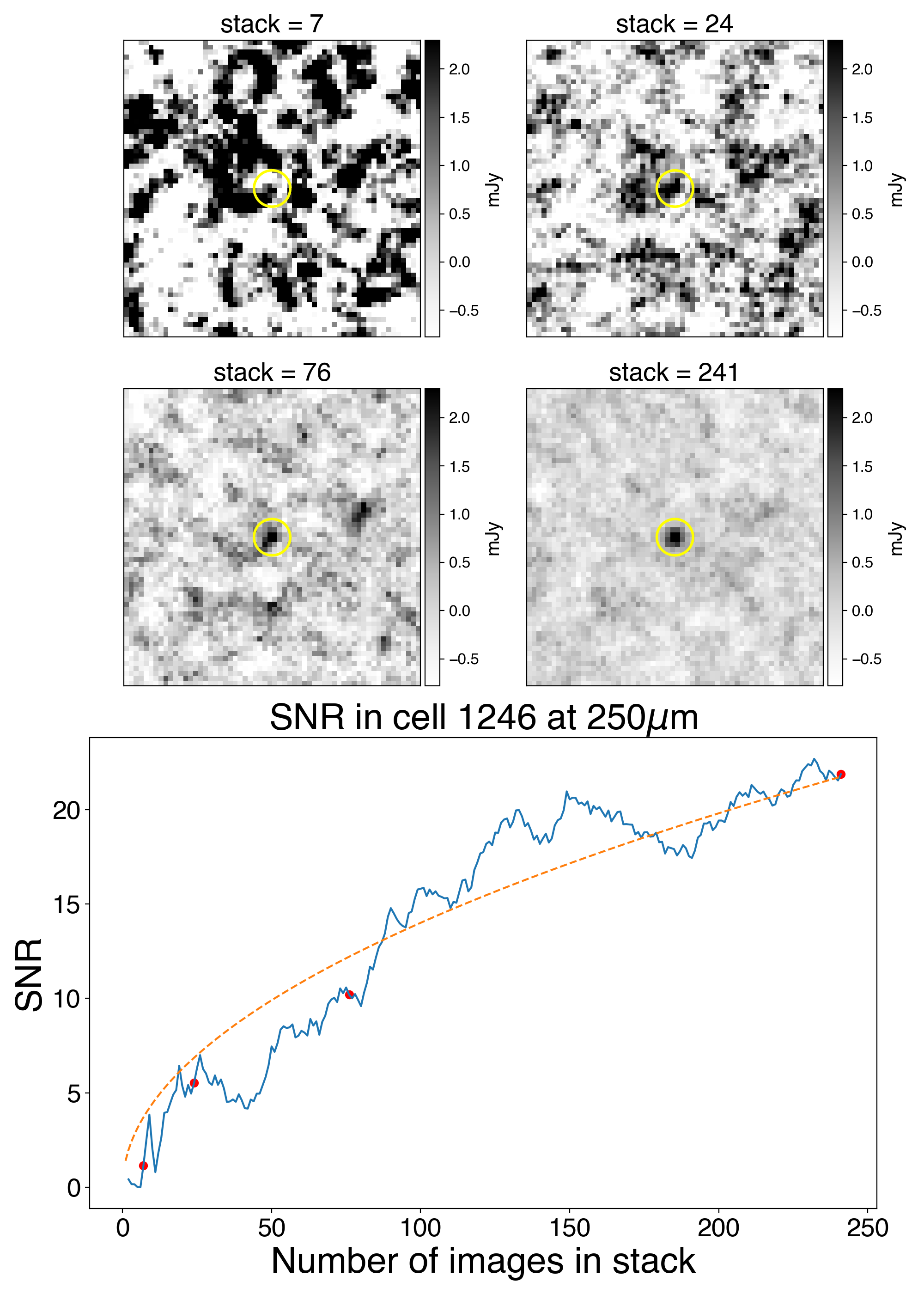}
\caption{  Example of the stacking process at 250$\mu$m. Top: Results of stacking 7, 24, 76, 241 images from the cell 1246.  Bottom: The SNR increase with the number of stacks, and roughly follows the trend of $\sqrt{N}$ for N numbers of images in stack (orange dashed line). 
\label{fig:stacking}}
\end{figure}

We demonstrate the stacking process of the 250$\mu$m images in cell 1246 as an example in Figure~\ref{fig:stacking}. There are 241 galaxies assigned to cell 1246 as the best-matching-cell.  On the top panel we show the progress of stacking when the number of images are logarithmically spacing at 7, 24, 76, 241 images. The background of the image progressively becomes smoother.  The signal-to-noise ratio (SNR) as a function of number of stacks is shown in the bottom panel, which roughly follows the square root of the number of stacks when the source is detected.

\subsection{SOM Mosaic of IR stacks}

\begin{figure}[h!]
\centering

\includegraphics[width=1\linewidth]{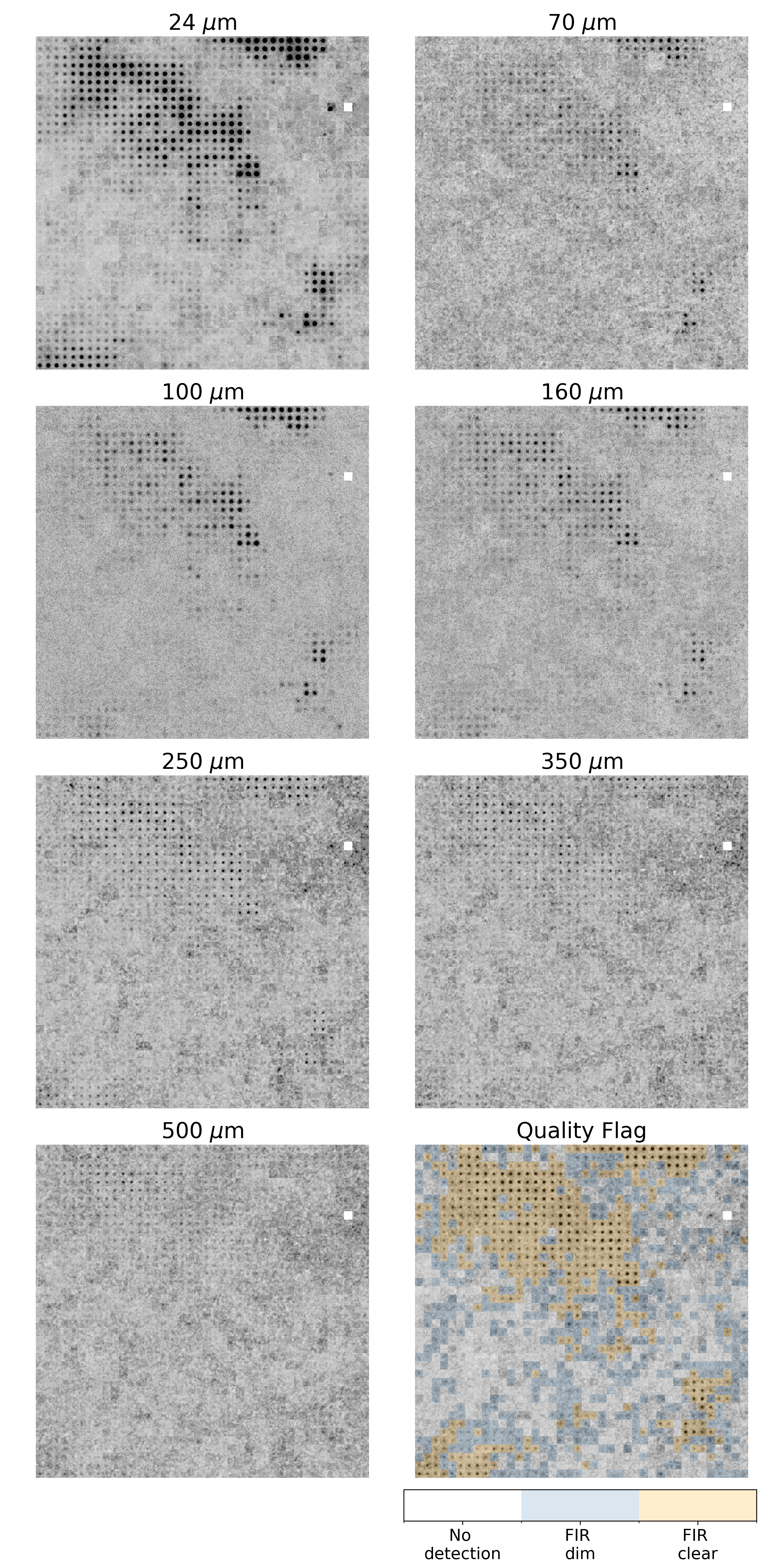}
\caption{ The mosaic of the stacked images for each cell at different IR wavelengths.  Each stacked image is normalized by its dispersion of the background before aligning to the mosaic.  The quality flags are plotted over the composite mosaic of 160$\mu$m, 250, 350, and 500 $\mu$m in the bottom right. 
\label{fig:SOM_mosaic}}
\end{figure}

To visually inspect the stacked images produced using our SOM, we create mosaic figures of each infrared band by chopping the stack images around the center after normalizing to the dispersion of the background, and aligning these chopped images according to their coordinates of the cells on the SOM.   
As shown in Figure~\ref{fig:SOM_mosaic}, cells with clear IR detections are close to each other on the SOM. 
The distribution patterns of cells with clear signals in the mosaic figures appear repeatedly in different wavelengths.  
We classified the cells with different quality flags; ``FIR clear'' if 4 or more IR bands from 24$\mu$m to 500$\mu$m are detected. ``FIR dim'' if 2-3 IR bands from 24$\mu$m to 500$\mu$m are detected. The rest of the cells are labeled as ``No detection''. We note that the ``No detection" cells may have 1 IR band detected, which is the 24$\mu$m that is not sufficient to study the FIR emission shape.  We plot the quality flags over the composite mosaic of 160$\mu$m, 250, 350, and 500 $\mu$m in the bottom right of Figure~\ref{fig:SOM_mosaic}.  
Among the 1600 cells in our SOM,  460 cells are flagged as ``FIR clear", which includes 37$\%$ of the sources. 516 cells are flagged as ``FIR dim'', which includes 31$\%$ of the sources. 624 cells are flagged as ``No detection'', accounting for 32$\%$ of the sources. 
In the next section, we will constrain the FIR luminosities of the cells using models with different numbers of free parameters based on the quality flags.

\subsection{Simulation of the Stacking Process}\label{sec:simulation}

In the stacking procedure of this work, we scale the galaxies' FIR images using the scaling factor in equation~\ref{eq:scaling_factor} that normalizes the optical-to-NIR magnitudes to the median value of galaxies in the same cell. When we predict the FIR photometry of an individual galaxy using these stacked FIR images, we will reverse the scaling that matches their optical-to-NIR magnitudes.  This method is based on the assumption that the FIR SED shape can be extrapolated from the optical-to-NIR SED shape.  In this assumption, galaxies with similar optical-to-NIR SED will have similar FIR SED.  The correlation between the optical-to-NIR and the FIR is expected, since the optical-to-NIR SED is affected by the dust and star formation, which are the main drivers of FIR emission \citep{daCunha_2008, Jespersen_2025}.

We perform a simulation with mock images to study how the validity of optical-to-NIR to FIR extrapolation impacts our SOM prediction. We use the Herschel 250$\mu$m images to demonstrate the simulated stacking process, while the effects in other wavelengths are the same. We first select cells in our SOM, and generate the same numbers of mock FIR images as the occupation numbers of the cells.  In this set of mock FIR images, we will assume the true FIR flux densities $f_{\nu}(gal)$ of the galaxies follow a log-normal distribution. We will then add the noise to the mock images so that only the brightest 10$\%$ of the galaxies are individually detected.

The mock images of the galaxies are simulated by scaling the PSF images to the FIR flux densities $f_{\nu}(gal)$ and add the Gaussian background noises. We determine the distribution of the FIR flux densities $f_{\nu}(gal)$ by introducing a random scatter $\sigma_{int}$ from a log-normal distribution $LogN(0, \sigma_{int})$ on top of the optical-to-NIR linear flux densities $f_{cell}(gal)$ in each SOM cells: 
\begin{equation} \label{eq:simulation_fnu}
 f_{\nu}(gal) \sim a_{90}\times(~f_{cell}(gal) \times {LogN(0, \sigma_{int}^2)}~) ,  
\end{equation}
where the constant $a_{90}$ scales the distribution so that the 90th$\%$ percentile brightest galaxies match similar SNR as the individually detected objects in Section~\ref{sec:photometry}.  
The log-normal scatter $\sigma_{int}$ represents the FIR intrinsic scatter when extrapolating the FIR emission from the optical-to-NIR SED, while the optical-to-NIR flux distributions in the cells are also roughly a log-normal distribution.   
When the scatter is 0 dex, the FIR flux density, $a_{90}\times f_{cell}(gal)$, is perfectly extrapolated from the optical-to-NIR SED. If the FIR intrinsic scatter is $>0.3$ dex, comparable to the scale of dispersion in the optical-to-NIR SED, the extrapolation validity will vanish.  
During the stacking process, we normalize the FIR images with Equation~\ref{eq:scaling_factor}. This process would make the FIR flux distribution more concentrated if the FIR intrinsic scatter is small ($<$ 0.3 dex), therefore making the stacking process more robust.   

We use this simulation to study (1) the accuracy of predicting individual galaxies' photometry using the stacked images in Section~\ref{sec:photometry}, (2) the difference between the noisy and noiseless stacked images under different stacking procedures in the Appendix~\ref{sec:bias_in_simulation}.

\subsection{Measuring the Far-Infrared Luminosity and Star-Formation Rate }

\begin{figure}
\includegraphics[width=0.99\linewidth]{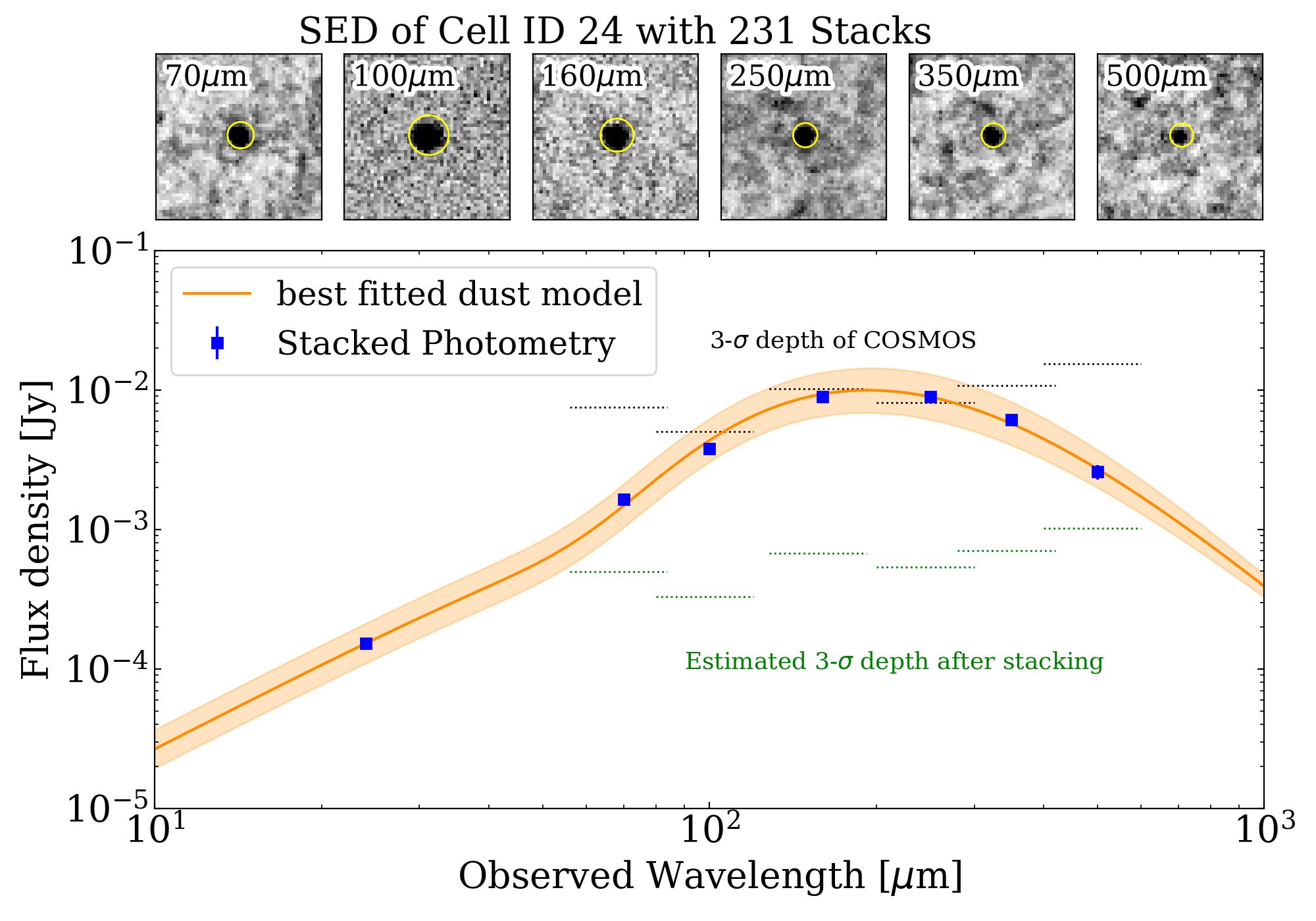}
\includegraphics[width=0.99\linewidth]{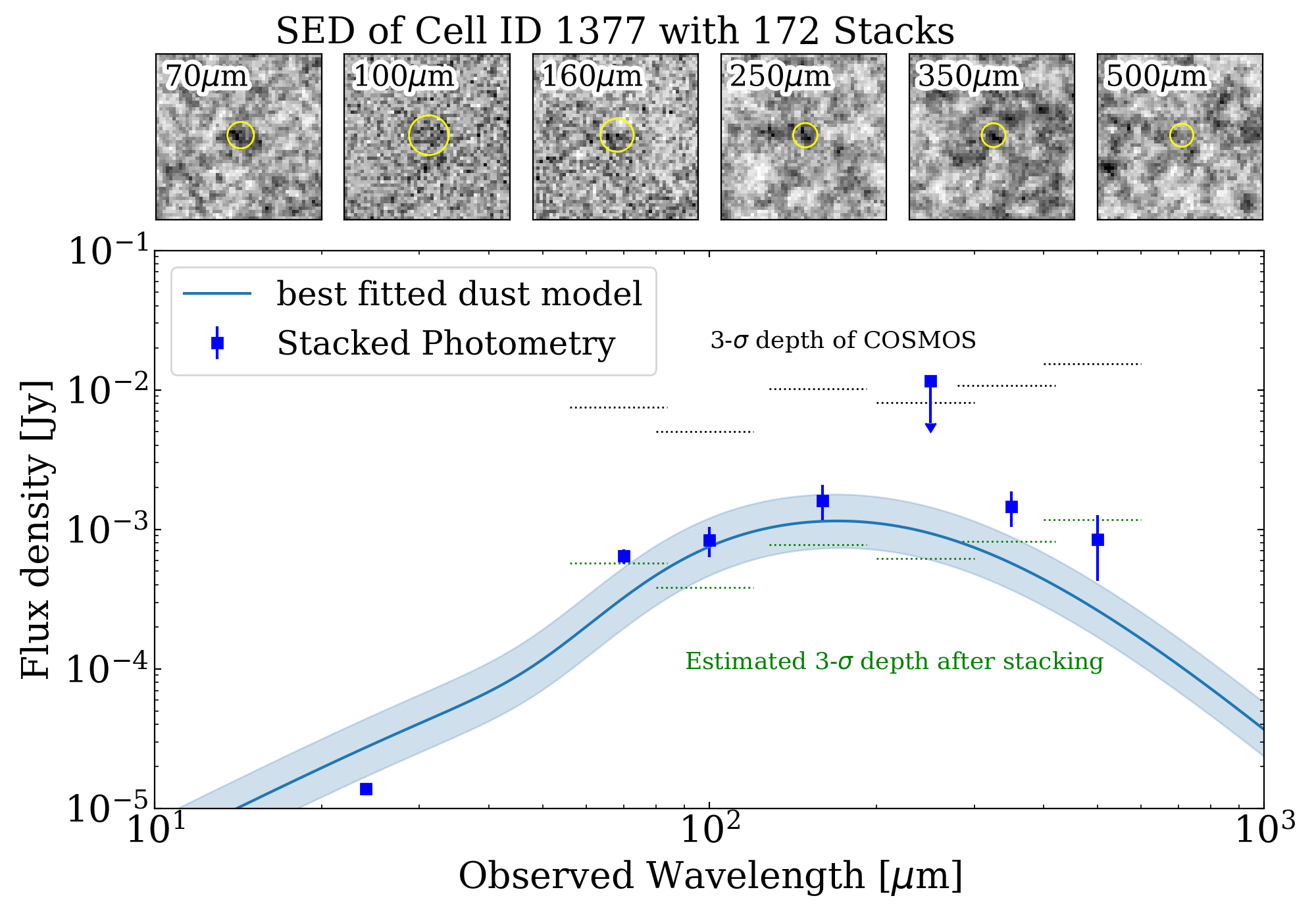}
\includegraphics[width=0.99\linewidth]{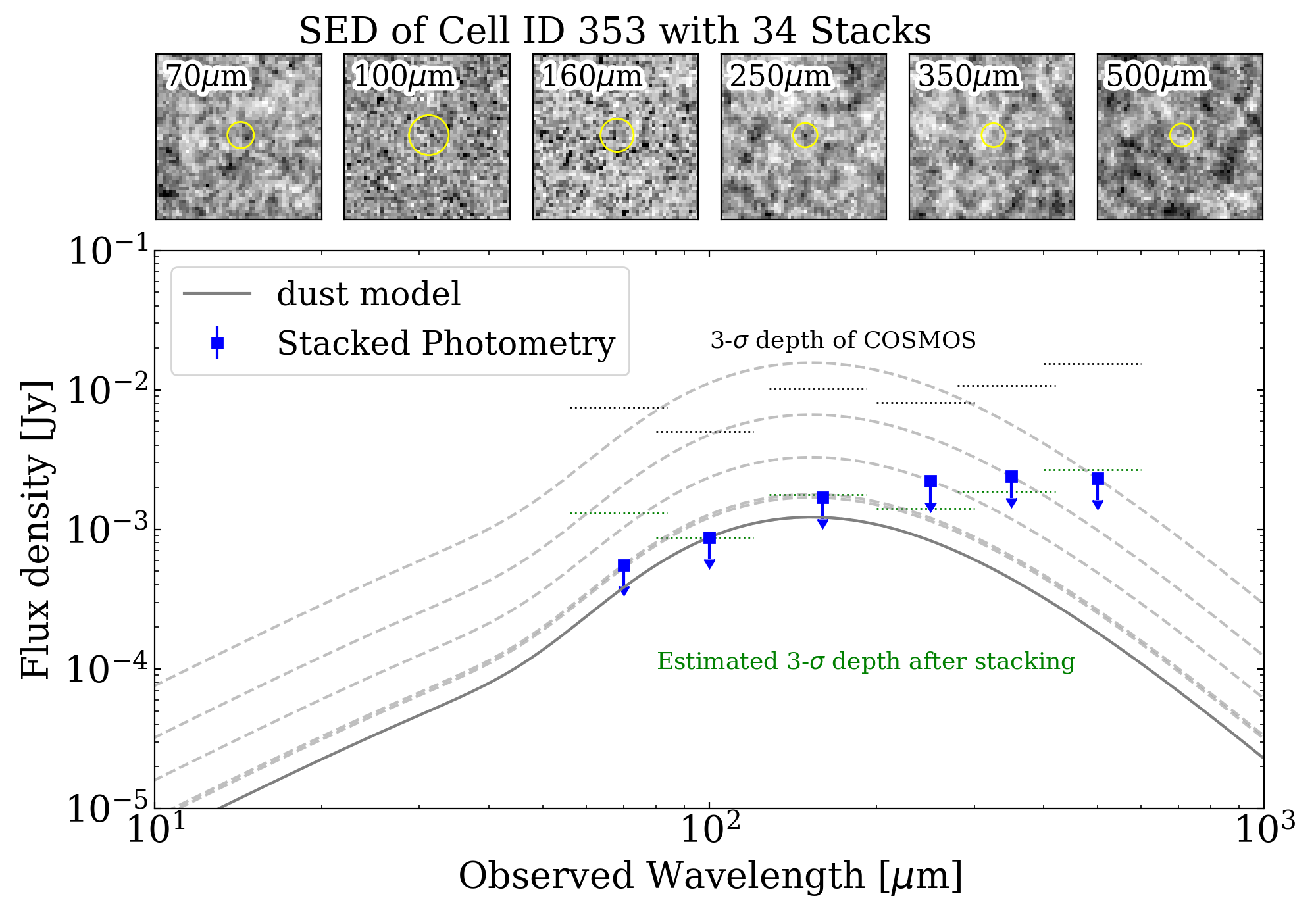}
\caption{  Top: The stacked SED of cell ID 24 as an example of the  ``FIR-clear'' labeled cell.  
Middle: The stacked SED of cell ID 1377 as an example of the  ``FIR-dim'' labeled cell. 
Bottom: The stacked SED of cell ID 353 as an example of the  ``no detection'' labeled cell.
\label{fig:FIR_SED_cells}}
\end{figure}

After the stacking the IR images and measuring the IR photometry of the cell representation in Section~\ref{sec:stacking}, we will estimate the FIR luminosity with the 24, 70, 100, 160, 250, 350, 500$\mu$m flux density. 
We fit the IR photometry with the dust emission model from \citet{Casey_2012}, described as the following: 
\begin{equation}
S_\nu{ \lambda } = N_{bb} \frac{ (1 - e^{-(\lambda_0/\lambda)^\beta})\lambda^{-3} }{e^{hc/k\lambda T}-1} + N_{pl}\lambda^{\alpha}e^{-(\lambda/\lambda_c)^2}.
\end{equation}
The first term in the model is a blackbody of cold dust at temperature $T$, modified by its opacity with optical depth $\tau(\lambda) = (\lambda_0/\lambda)^{\beta}$, where $\lambda\equiv 200\mu m$ is the wavelength where opacity is unity. 
The second term is a power law that represent a combination of warmer dust sub-components or dust heated by AGN. The power-law turnover wavelength $\lambda_c$ depends on the power-law slope $\alpha$ and the cold dust temperature $T$:  
\begin{equation}
\lambda_c \equiv \frac{3}{4} [ (b_1 + b_2 \alpha)^{-2} + (b_3 + b_4 \alpha)\times T ]^{-1}, 
\end{equation}
where $b_1 = 26.68$, $b_2 = 6.246$, $b_3 = 1.905\times10^{-3}$, and $b_4 = 7.243\times10{-5}$. 
The power-law normalization constant $N_{pl}$ scales with the blackbody normalization constant $N_{bb}$ and the turnover wavelength $\lambda_c$ in order to generate smooth transition between MIR and FIR: 
\begin{equation}
N_{pl} = N_{bb} \frac{ (1 - e^{-(\lambda_0/\lambda_c)^\beta})\lambda_c^{-3} }{e^{hc/k\lambda_c T}-1} .
\end{equation}
The dust emission model has up to four free parameters: $N_{bb}$, $T$, $\alpha$, and $\beta$.  
Empirically, the emissivity $\beta$ varies from 1 to 2.5, and the power-law slope $\alpha$ ranges from 1.5 (flatter slope; more warm dust) to 2.5 (steep slope; less warm dust) \citep{Dale_2001}. 
We perform a Markov Chain Monte Carlo approach to fit the above model to the FIR photometry. 
We fix the redshift using the $z_{photo}$ median inferred from the optical-to-NIR colors (Section~\ref{sec:data}), since the FIR photometry alone is not sufficient to constrain redshifts due to the degeneracy with the dust temperature and dust opacity.  
All 4 parameters are used if the cells are flagged as ``FIR clear" (having at least 4 FIR bands detected). We show an example of a `FIR clear" cell in the top panel of Figure~\ref{fig:FIR_SED_cells}.   For cells flagged as ``FIR dim" (2-3 FIR bands detected), such as the example in the middle panel of Figure~\ref{fig:FIR_SED_cells}, we fit the FIR spectrum with fixed $\alpha=2$ and $\beta=1.6$ that is empirically used \citep{Casey_2012}.  Cells with no FIR bands detected are fitted using the 3$\sigma$ flux uncertainty with the dust emission model fixed at $T=40K$, $\alpha=2$, and $\beta=1.6$. The dust temperature $T$ (in specific, the dust {\it SED} temperature) is typically $5\sim15$K higher than the dust peak temperature, defined as $T_{peak} \equiv$ b/$\lambda_{peak}$, where $\lambda_{peak}$ is the wavelength at the peak of the FIR SED.  We choose this dust temperature of 40K based on the observed redshift evolution of the $T_{peak}$ \citep{Faisst_2020}. As shown in the bottom panel of Figure~\ref{fig:FIR_SED_cells}, we calculate the upper limit of the FIR SED using each band independently, since the upper limits of each band are not necessarily correlated.   We then integrated the modeled FIR spectra from rest-frame 8$\mu$m to 1000$\mu$m to obtain the total IR luminosity, $L_{IR}$.

We then derive the FIR star-formation rates using the Kennicutt relation \citep{K98}, assuming the Chabrier initial mass function \citep{Chabrier_2003}: 
\begin{equation}
{\rm SFR}_{\rm FIR}~[M_\odot~yr^{-1}]= 2.8\times10^{-44}\times L_{\rm IR}~[erg~s^{-1}].
\end{equation}
The FIR luminosity of the individual galaxies within SOM cells will be predicted by multiplying $L_{\rm IR}$ over the scaling factor $S$ from equation~\ref{eq:scaling_factor}. 
The summarized table of labeled features and measurements of the SOM, as well as the table of derived properties for individual galaxies are available in: ({\it link will be added upon publication of this paper}).

%%%%%%%%%%%%%%%%%%%%%%%%%%%%%%%%%%%%%%%%%%%%%%%%%%%%%%%%%%%%%%%%%%
%%%%%%%%%%%%%%%%%%%%%%%%%%%%%%%%%%%%%%%%%%%%%%%%%%%%%%%%%%%%%%%%%%
\section{Results and Discussions} \label{sec:result}
%%%%%%%%%%%%%%%%%%%%%%%%%%%%%%%%%%%%%%%%%%%%%%%%%%%%%%%%%%%%%%%%%%
%%%%%%%%%%%%%%%%%%%%%%%%%%%%%%%%%%%%%%%%%%%%%%%%%%%%%%%%%%%%%%%%%%

\subsection{The cells with FIR Detection }\label{sec:cell-label}
 
\begin{figure}
\includegraphics[width=0.99\linewidth]{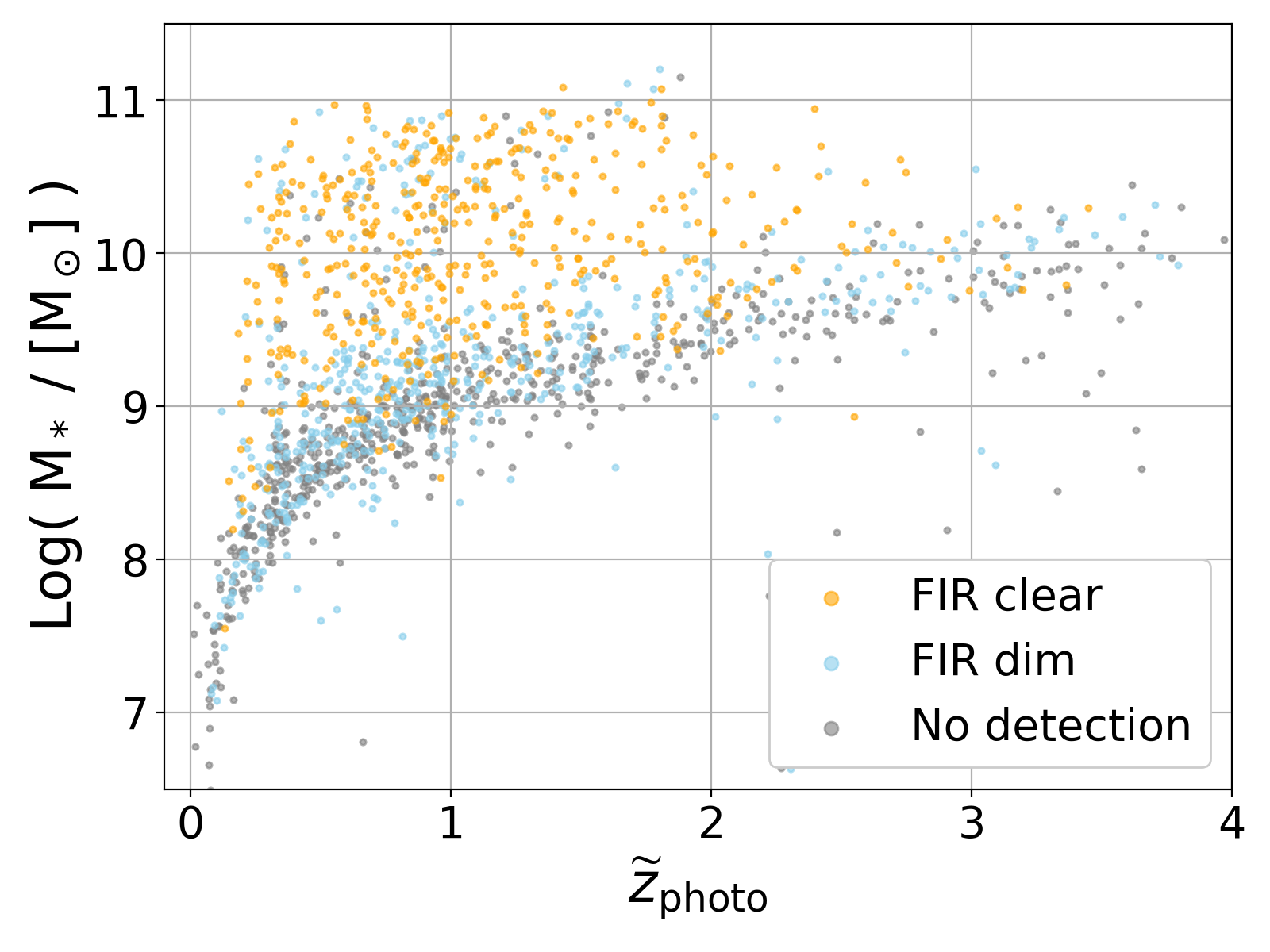}
\caption{  The median values of Redshift vs stellar mass of galaxies from COSMOS2020 assigned to the SOM cells, color-coded with the FIR quality flags. 
\label{fig:flag_scatter1}}
\end{figure}

In Figure~\ref{fig:flag_scatter1} we show the median stellar mass and redshift \mzph\ of the galaxies assigned to the SOM cells, color-coded with the FIR quality flags. 
The FIR images of cells with ``FIR clear'' quality flags are assigned to galaxies that have sufficiently high stellar mass at the given redshift, so that their FIR signals are above detection limits after the enhancement from stacking.  

\begin{figure*}
\includegraphics[width=0.99\linewidth]{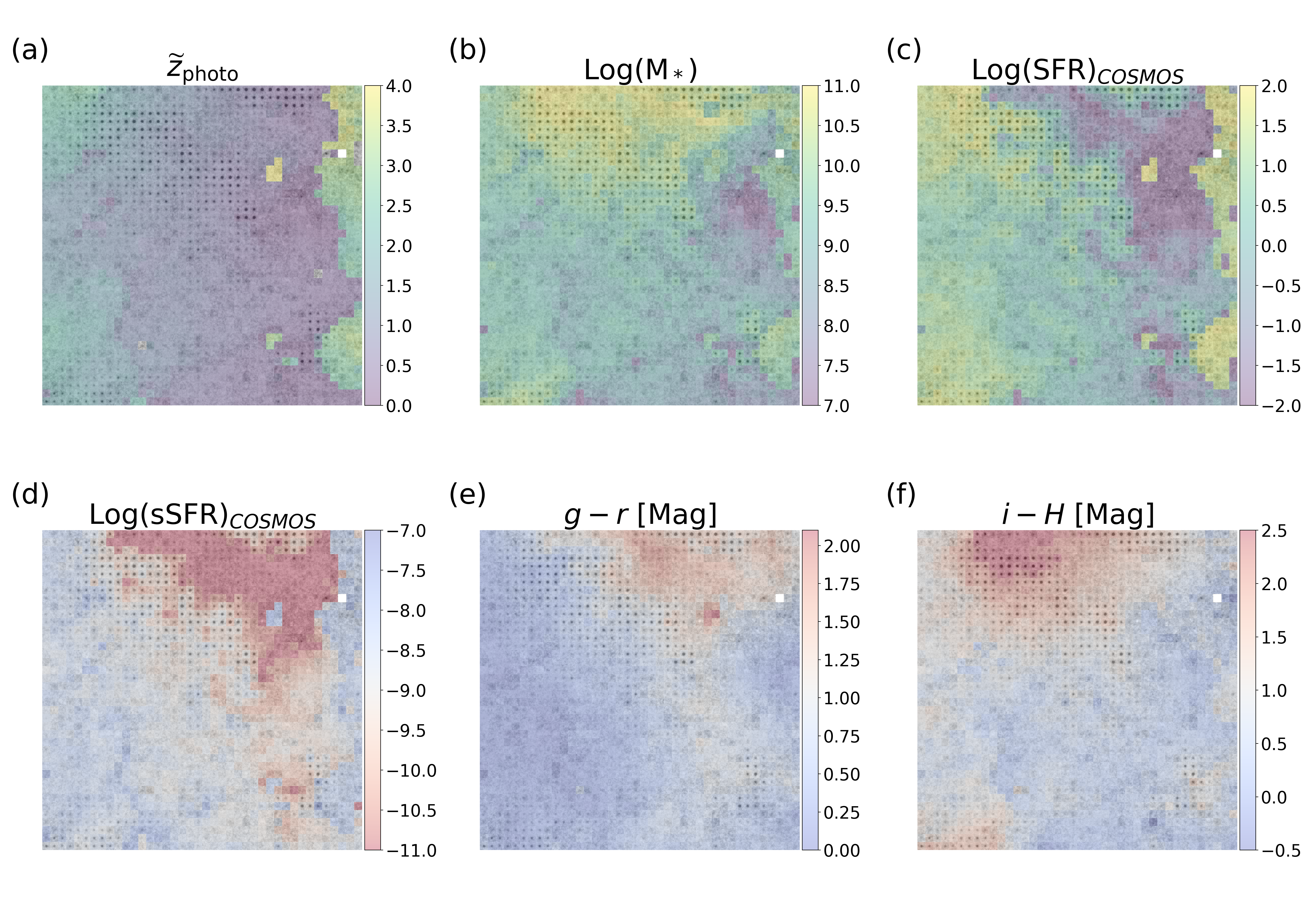}
\caption{  The stack image mosaic composited from 160$\mu$m, 250$\mu$m, 350$\mu$m, and 500$\mu$m, plot over the SOM labeled with median values of (a) redshifts, (b) Log($M_*$), (c) Log(SFR) (d) Log(sSFR) (e) $g - r$ (f) $i - H$. The Log(SFR) and Log(sSFR) shown in this figure are derived from the Optical-to-NIR SED in the COSMOS2020 catalog. 
\label{fig:map_and_mosaic}}
\end{figure*}

In Figure~\ref{fig:map_and_mosaic} we over-plot the stack image mosaic composited from 160$\mu$m, 250$\mu$m, 350$\mu$m, and 500$\mu$m on the SOM labeled with the median values of the assigned galaxies' \mzph, stellar mass, SFR, sSFR, the $g-r$ and the $i-H$ colors. The stellar mass, SFR and sSFR are adopted from the optical-to-NIR SED in the COSMOS2020 \citep{Weaver_2022}.  There is no strong dependence of FIR signals and \mzph\ alone, while FIR of cells labeled with \mzph$>3$ are mostly undetected.  The galaxy M$_*$ on the ``FIR clear" cells are moderate to high stellar mass ($>10^8 M_\odot$). These galaxies are mostly the star-forming main sequence galaxies, whose SFR scales with the stellar mass.  In the sSFR panel, we see that a high sSFR does not guarantee FIR detection, since most of the high sSFR (starburst) galaxies are low mass galaxies ($<10^8 M_\odot$). However, low sSFR galaxies (which are more quiescent) are consistently not detected in the FIR, which is because of their lower SFR as well as lower abundance of dust in general.
Finally, the NIR red color ($i-H$) shows stronger similarity to the FIR detection pattern than the optical red color ($g-r$). 

\begin{figure*}
\includegraphics[width=0.99\linewidth]{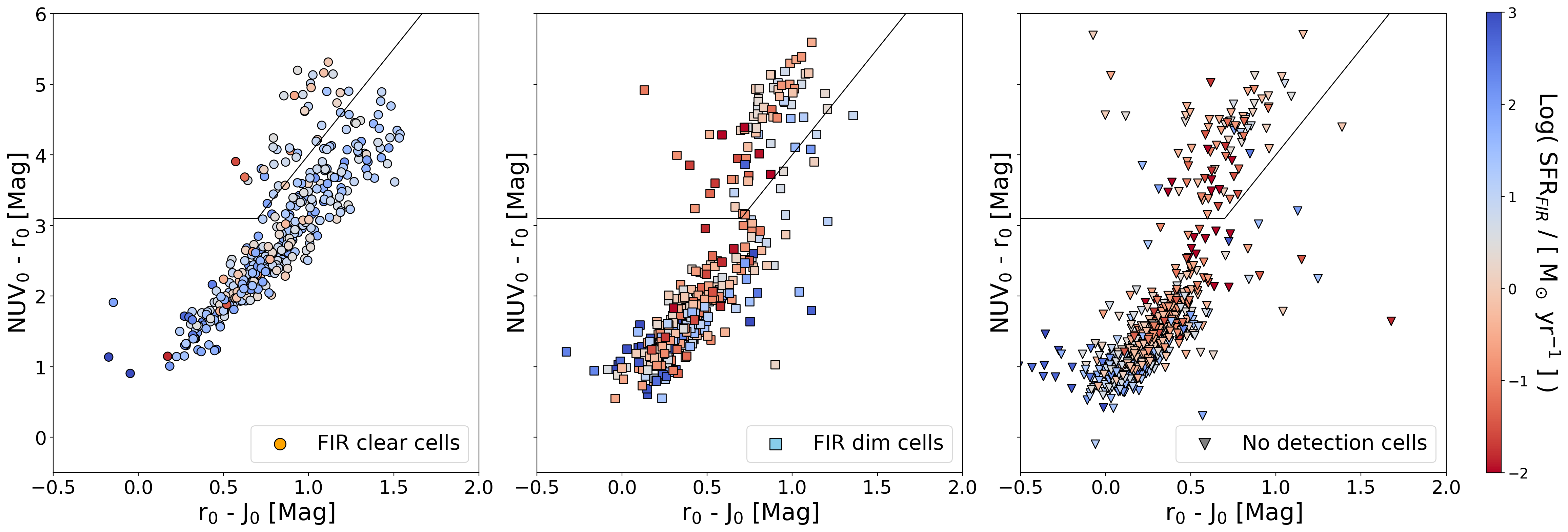}
\caption{  The distribution of the cells flagged as (Left) ``FIR clear", (Middle) ``FIR clear", and (Right)``No emission" in the rest-frame $NUV_0-r_0$ vs $r_0-J_0$ diagram, color-coded with Log(SFR$_{FIR}$). The SFR$_{FIR}$ labeled on the ``No emission" cells are upper limits. 
\label{fig:NUVrJ}}
\end{figure*}

We also show the ``$NUVrJ$" diagram in Figure~\ref{fig:NUVrJ} as a different perspective.  Galaxies distributed in the upper-left corner of the $NUVrJ$ diagram are classified as quiescent galaxies, and galaxies distributed outside the corner are classified as star-forming galaxies \citep{Ilbert_2013}.  The ``$NUVrJ$" classification is well in agreement with the sSFR \citep{Faisst_2017}.  We labeled the SOM cells with the median values of the assigned galaxies' rest-frame magnitudes NUV$_0$, $r_0$, and $J_0$  adopted from COSMOS2020.  In the left panel of Figure~\ref{fig:NUVrJ}, the ``FIR clear" cells distribute in both sections of star-forming galaxies and quiescent galaxies.  The few cells of quiescent galaxies that have FIR detection are distributed at the edge of the ``low sSFR valley" on the SOM in Figure~\ref{fig:flag_scatter1}, labeled with high stellar mass.

\subsection{Predicting individual galaxy photometry from the calibrated SOM}\label{sec:photometry}

The FIR photometry and SFR of the individual galaxies are calculated by multiplying these properties of their best-matching cell by the galaxies' scaling factors based on the optical-to-NIR magnitude difference in the equation~\ref{eq:scaling_factor}. By doing so, we are assuming that the galaxies with similar SED in the optical and NIR also have similar SED in the FIR, and the FIR to optical-to-NIR relation and be scaled up and down linearly within a cell.  

\begin{figure}
\includegraphics[width=0.99\linewidth]{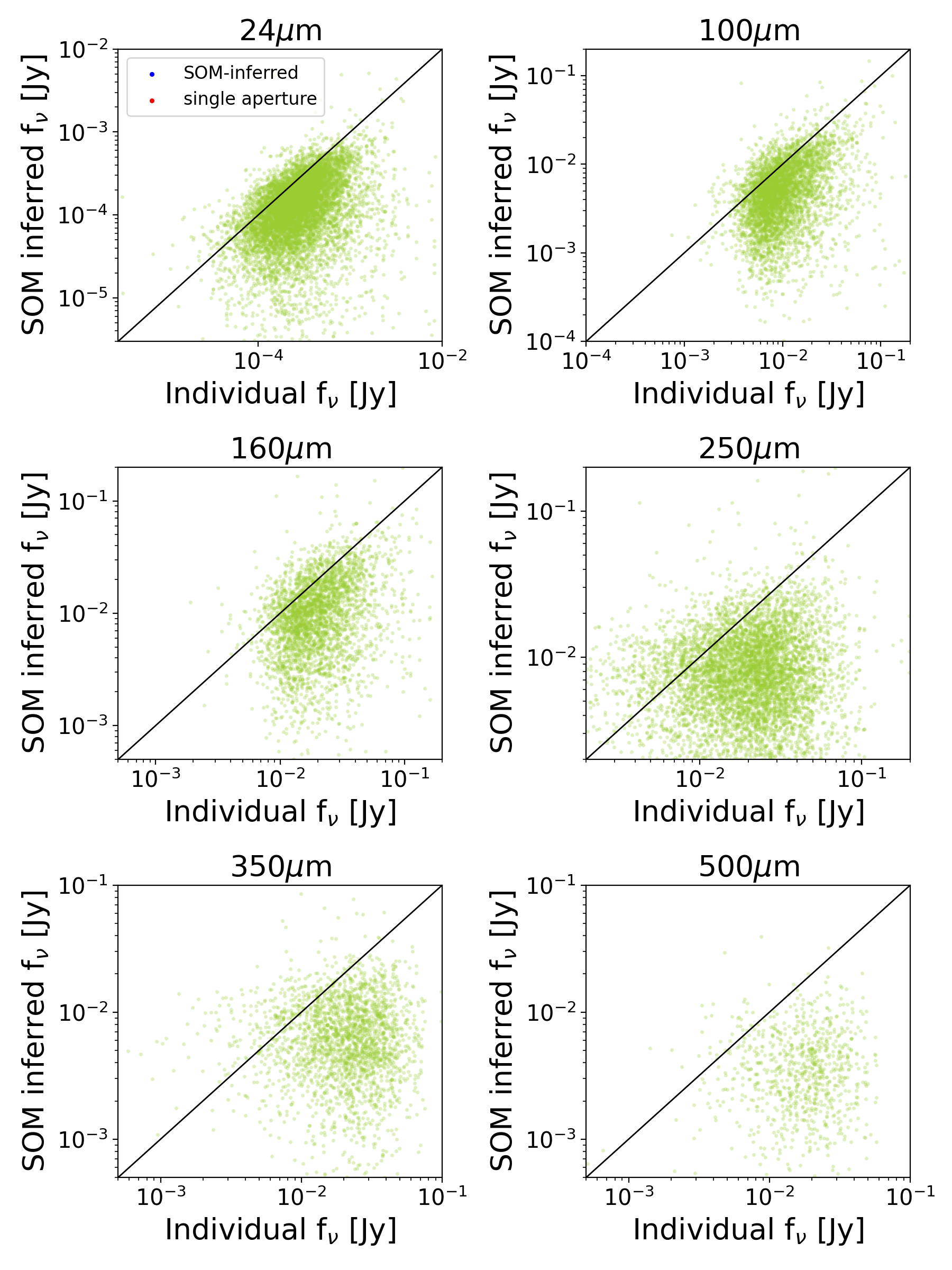}
\caption{ The flux density of individual FIR-bright galaxies selected in \citet{Jin_2018} measured directly v.s. the predicted using our SOM in Spitzer MIPS 24$\mu$m, Herschel PACS 100$\mu$m, 160$\mu$m, and Herschel SPHIRE 250$\mu$m, 350$\mu$m, 500$\mu$m.   The black line is the identical line. 
\label{fig:mag_measurement}} 
\end{figure}

We may verify our prediction on the FIR-bright galaxies where their FIR photometry are both detected in the single object images and the stacked image of the best-matching-cell in our SOM, with a caveat that these FIR-bright galaxies are all  distributed at the bright-end within their best-matching-cell. 
We match FIR-bright galaxies from the FIR catalogs from \citet{Jin_2018} with the coordinates, and in each band we require the signal-to-noise ratio greater than 5, and their best-matching-cell quality flag to be ``FIR-clear".   
Our FIR photometry comparisons are shown in Figure~\ref{fig:mag_measurement}.  
The SOM predicted photometry tend to underestimate the flux density as compared to their direct image measurements, and the dispersion are higher in the longer wavelengths.   We note that the data distribution appears much wider in the lower triangle than the upper triangle due to the asymmetry of the log-log scale in the figure. 

\begin{figure}
\includegraphics[width=0.99\linewidth]{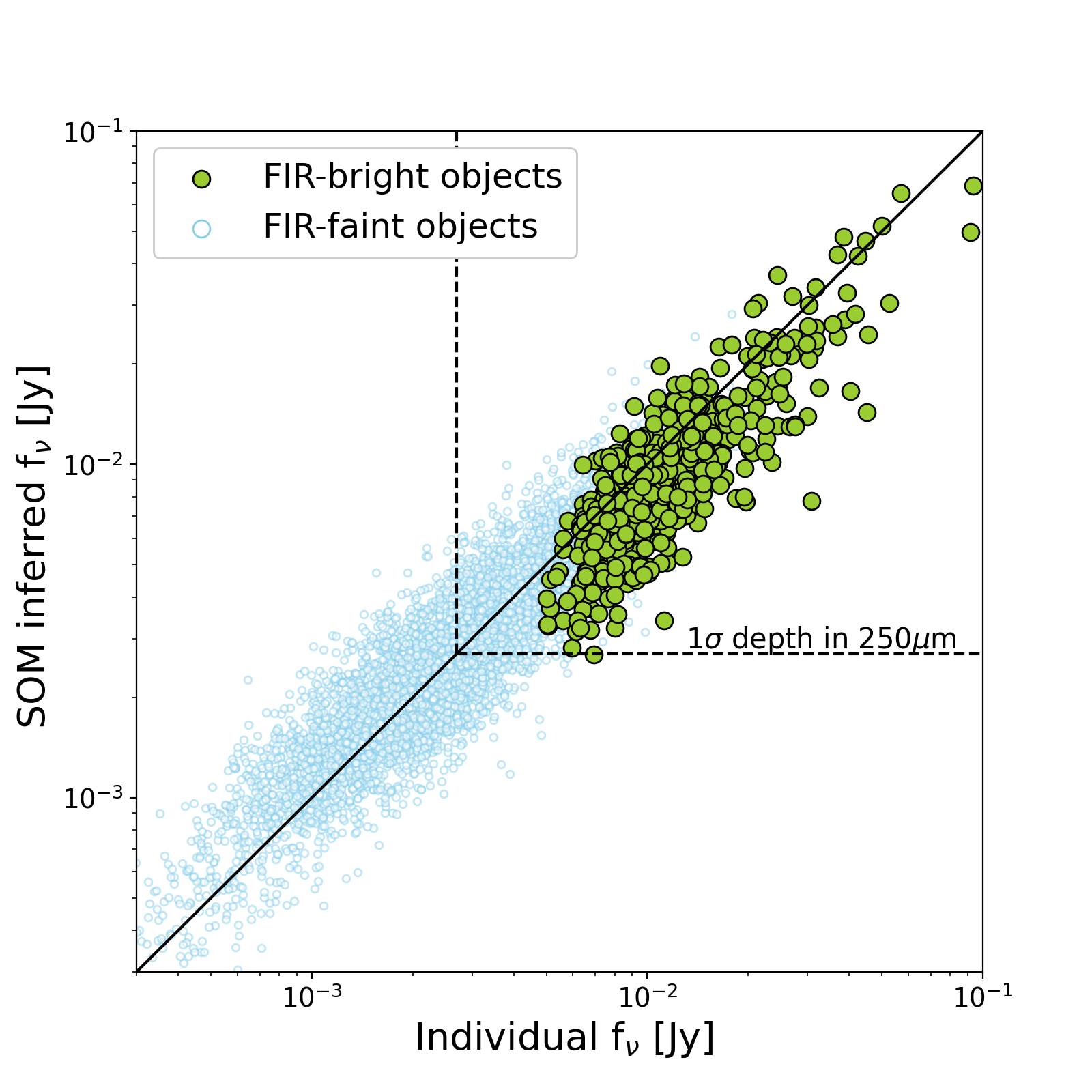}
\caption{  The individually measured flux density of simulated galaxies compare to the prediction using our SOM.   The black line is the identical line. The yellow-green points simulates the FIR-bright galaxies that can be detected individually, and the blue circles are the FIR-faint galaxies that we aim to predict using the SOM. 
\label{fig:model_measurement}} 
\end{figure}

At first glance, it seems like the SOM prediction underestimate the FIR photometry for all galaxies. However, the underestimation is mainly caused by predicting a biased sample with the median flux.   
We use the simulation described in Section~\ref{sec:simulation} to compare the input to SOM-predicted result of the  mock galaxies.  The intrinsic flux densities of the mock galaxies are simulated based the equation~\ref{eq:simulation_fnu}, where the FIR flux distribution of galaxies in a cell is similar to their optical-to-NIR flux  distribution, with an additional Lognormal scatter. The 90th$\%$ brightest galaxies represent the FIR-bright galaxies that can be detected individually.  
In Figure~\ref{fig:model_measurement} we show the simulated results, assuming the FIR intrinsic scatter $\sigma_{int} = 0.15$ in the equation~\ref{eq:simulation_fnu}. The SOM prediction will underestimate most of the FIR-bright objects, similar to our measurements presented in Figure~\ref{fig:mag_measurement}, while the entire distribution is not as biased as the bright sample (see~Appendix~\ref{sec:bias_in_simulation} for more discussion).  

\begin{figure}
\includegraphics[width=0.99\linewidth]{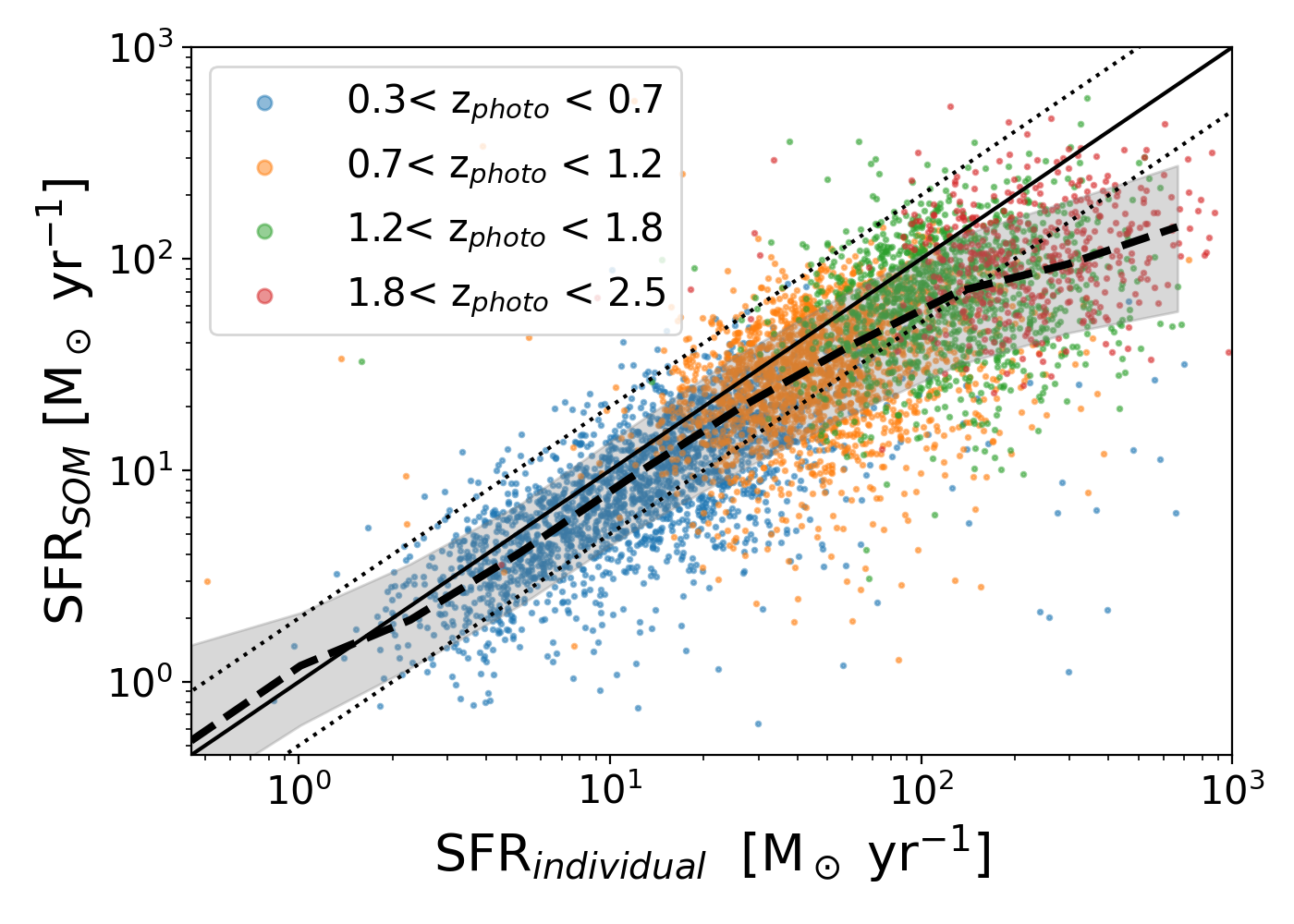}
\caption{ The individually measured SFR vs the SOM-inferred SFR of FIR-bright galaxies selected from \citet{Jin_2018}, color-coded in different redshift range. The black solid line marks the identical line, and the black dotted lines are $\pm0.3$ dex from the identical line.   The black dashed line indicates the median of the distribution, and the shadow area marks the 25$\%$ to 75$\%$ percentile.   
\label{fig:SFR_v_SFR}} 
\end{figure}

We also underestimate the SFR$_{FIR}$ with the SOM for these single-image detected galaxies, since the SFR is linear to the integrated FIR luminosity. In Figure~\ref{fig:SFR_v_SFR} we show that the SOM-prediction increasingly underestimates the SFR at higher SFR values.  
However, as we mentioned above, the underestimation occurs because these single-image detected galaxies are bias at the FIR-bright end compared to other galaxies assigned to the same best-matching cells. Their SFR are higher than most of the galaxies at the given stellar mass range, as we can see in Figure~\ref{fig:M_SFR}.

\begin{figure}
\includegraphics[width=0.99\linewidth]{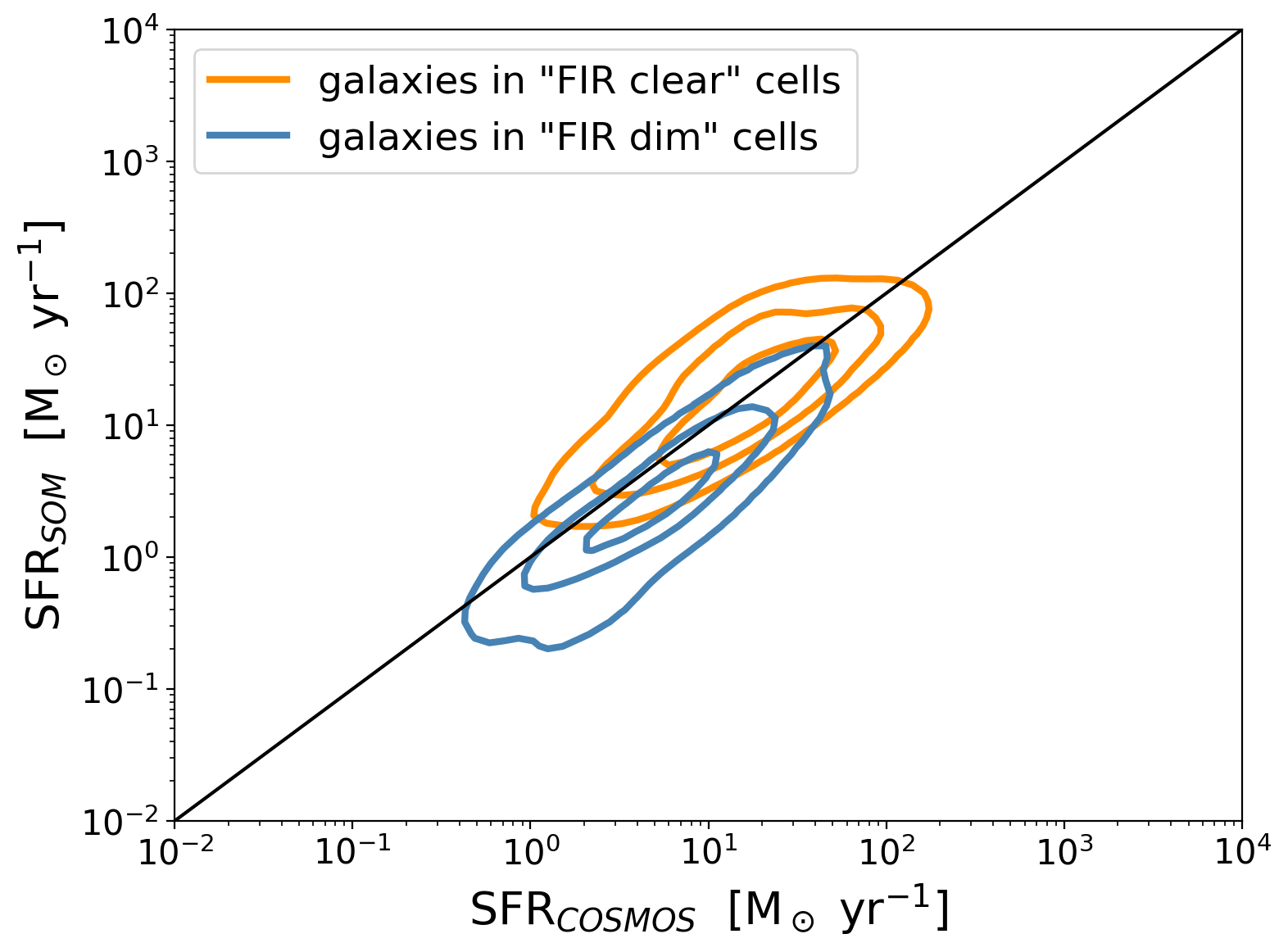}
\caption{ The SFR adopted in COSMOS2020 classic vs the SOM-inferred SFR for all the galaxies that are assigned to ``FIR clear" cells (orange contour) and ``FIR dim" cells (blue contour). The black solid line marks the identical line.  The outermost contours contain 99$\%$ of the data.   
\label{fig:SFR_v_SFR_cosmos}} 
\end{figure}

Although we cannot compare the individual FIR measurement for all the galaxies, we can compare our SFR$_{\rm FIR}$ predicted with our SOM to the SFR derived from optical-to-NIR template fitting in the COSMOS2020 classic catalog.  In Figure~\ref{fig:SFR_v_SFR_cosmos}, we show the contour distributions of galaxies assigned to ``FIR clear" and ``FIR dim" cells.  The Log(SFR$_{\rm FIR}$/SFR$_{\rm COSMOS}$)=$0.12^{+0.61}_{-0.41}$ for galaxies in the ``FIR clear" cells, and Log(SFR$_{\rm FIR}$/SFR$_{\rm COSMOS}$)=$-0.20^{+0.64}_{-0.48}$ for galaxies in the ``FIR clear" cells.

\subsection{The Star-Formation Rate - Stellar Mass Relation}\label{sec:SFR-M_relation}

The SFR-M relation is a tight relation regulating the star-forming galaxies, with a modest intrinsic scatter (~0.2–0.3 dex) that persists from the local Universe to high redshift z$\sim$6 \citep{noeske_2007,daddi_2007,khusanova_2021,clarke_2024}. The star-forming galaxies that follows the SFR-M relation are called the main sequence galaxies. 
The SFR-M relation can be described in this format \citep{Popesso_2023}: 
\begin{equation}\label{eq:SFR-M}
\text{log}SFR = a_0 + a_1 t - \text{log}(1 + (\frac{M_*}{M_{0}(t)})^{-\gamma}),
\end{equation}
where $a_0$ is a normalization constant, $a_1 <0$ varies the SFR by the age of the universe $t$ in Gyr, $M_{0}(t) = 10^{a_2 + a_3 t} ~M_{\odot}$ is the turnover mass at the given time. At low stellar mass where $M << M_0$, the SFR scales with the stellar mass as a power law with the slope of $\gamma \sim 1$. Therefore, the star-forming galaxies with M$_\star$ below the turnover mass $M_0$ exhibit an almost constant specific SFR. As the stellar mass increases, the SFR-M relation is flatten, and the sSFR is progressively suppressed.

\begin{figure*}[h]
\includegraphics[width=0.49\linewidth]{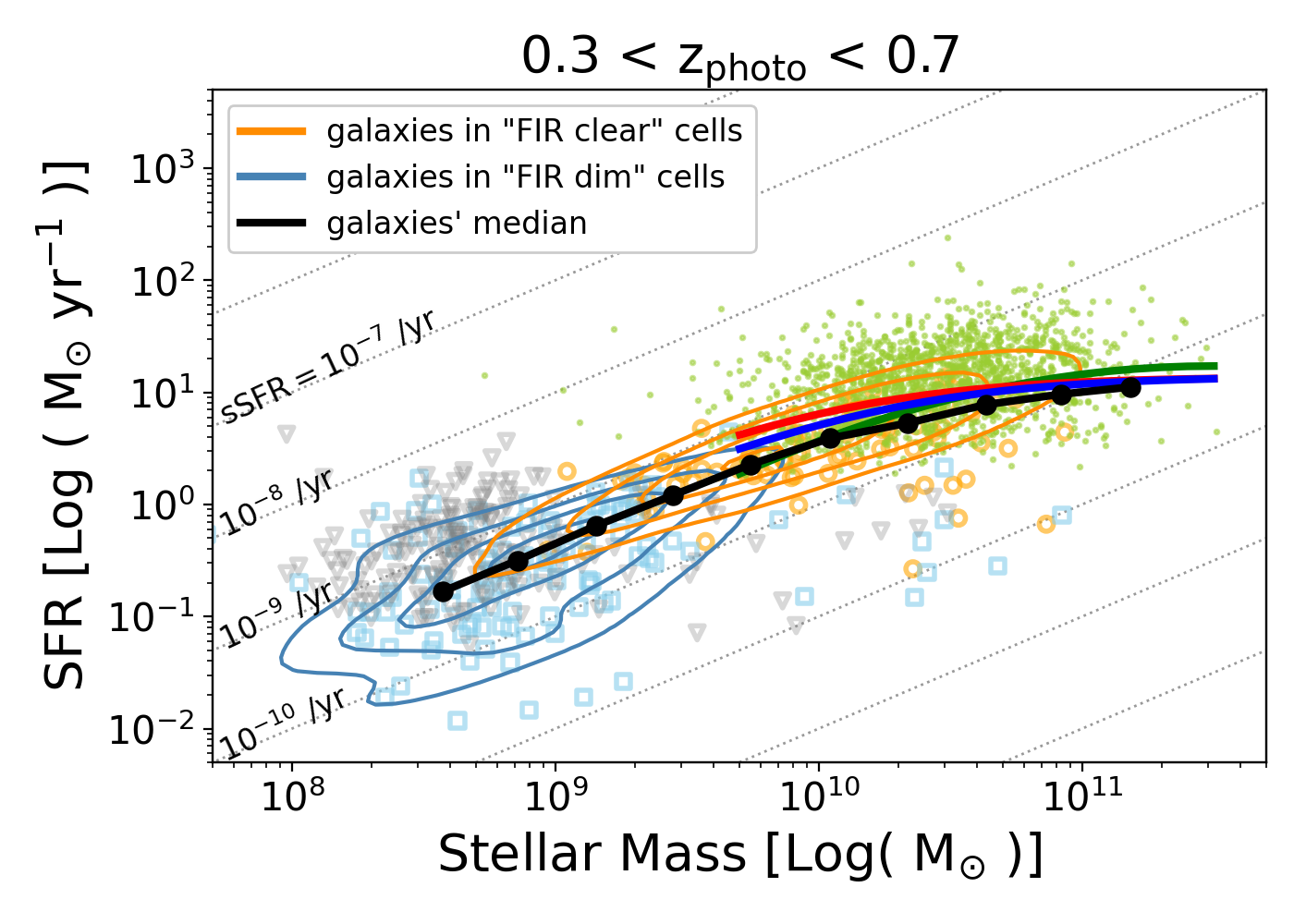}
\includegraphics[width=0.49\linewidth]{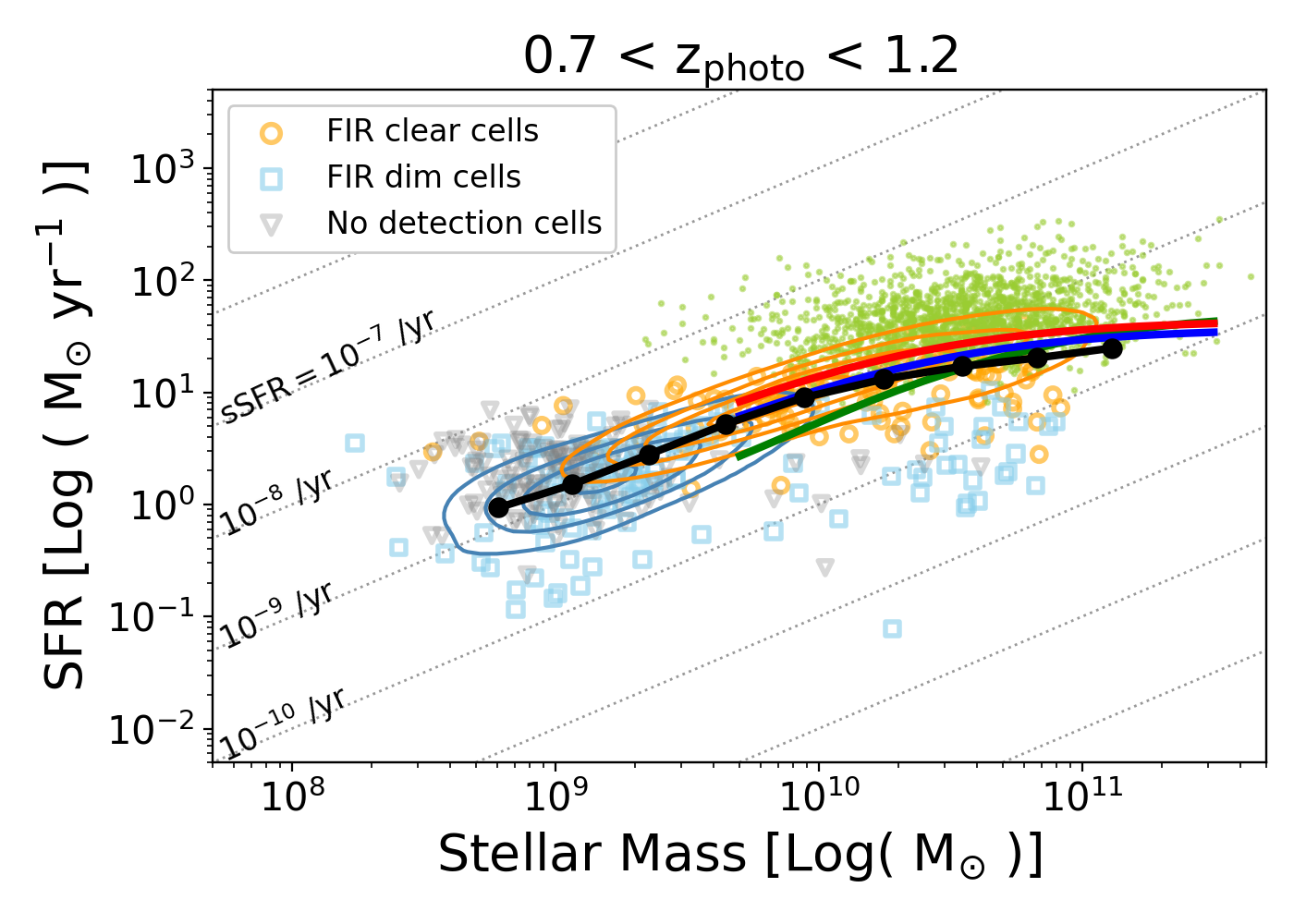}
\includegraphics[width=0.49\linewidth]{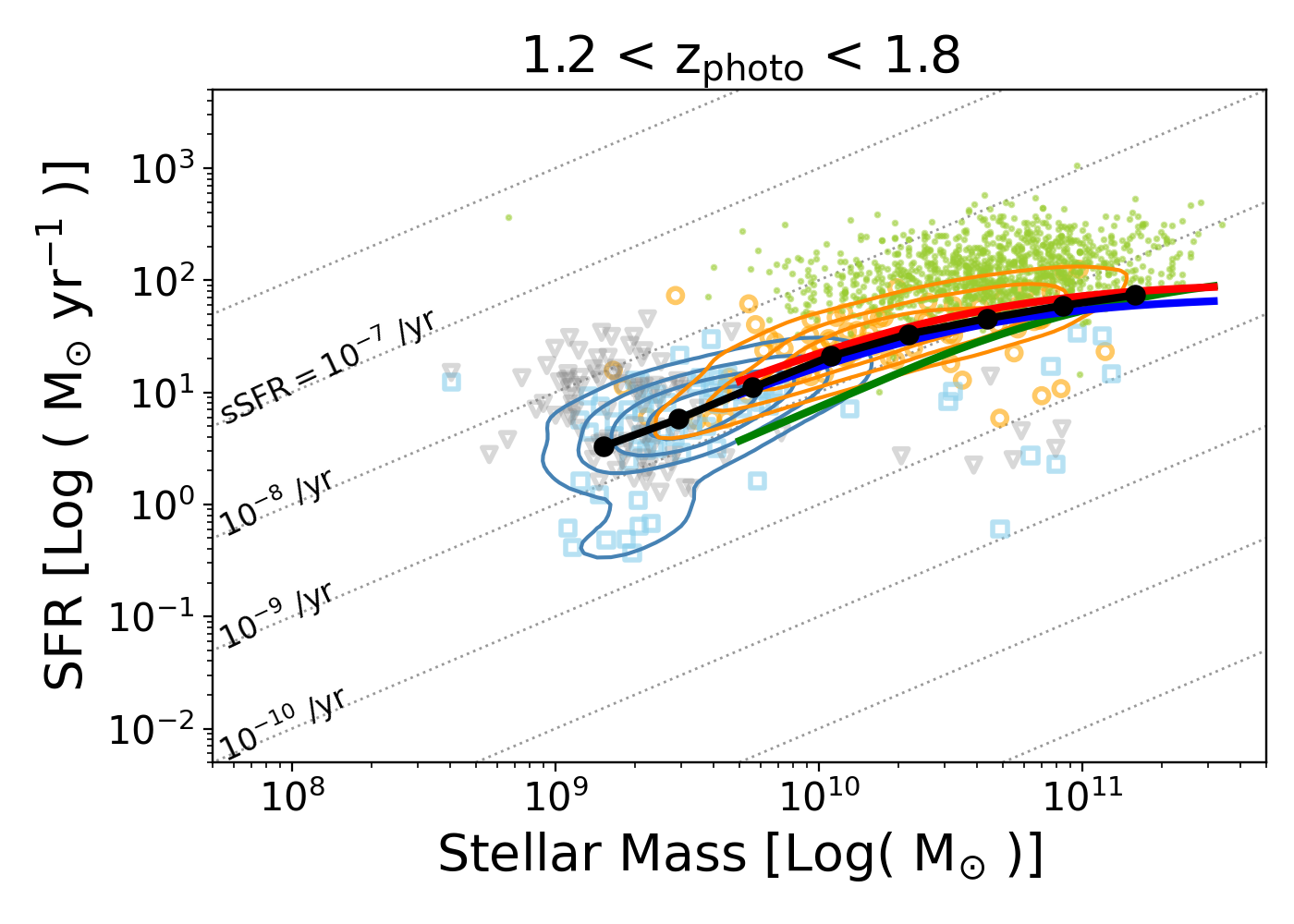}
\includegraphics[width=0.49\linewidth]{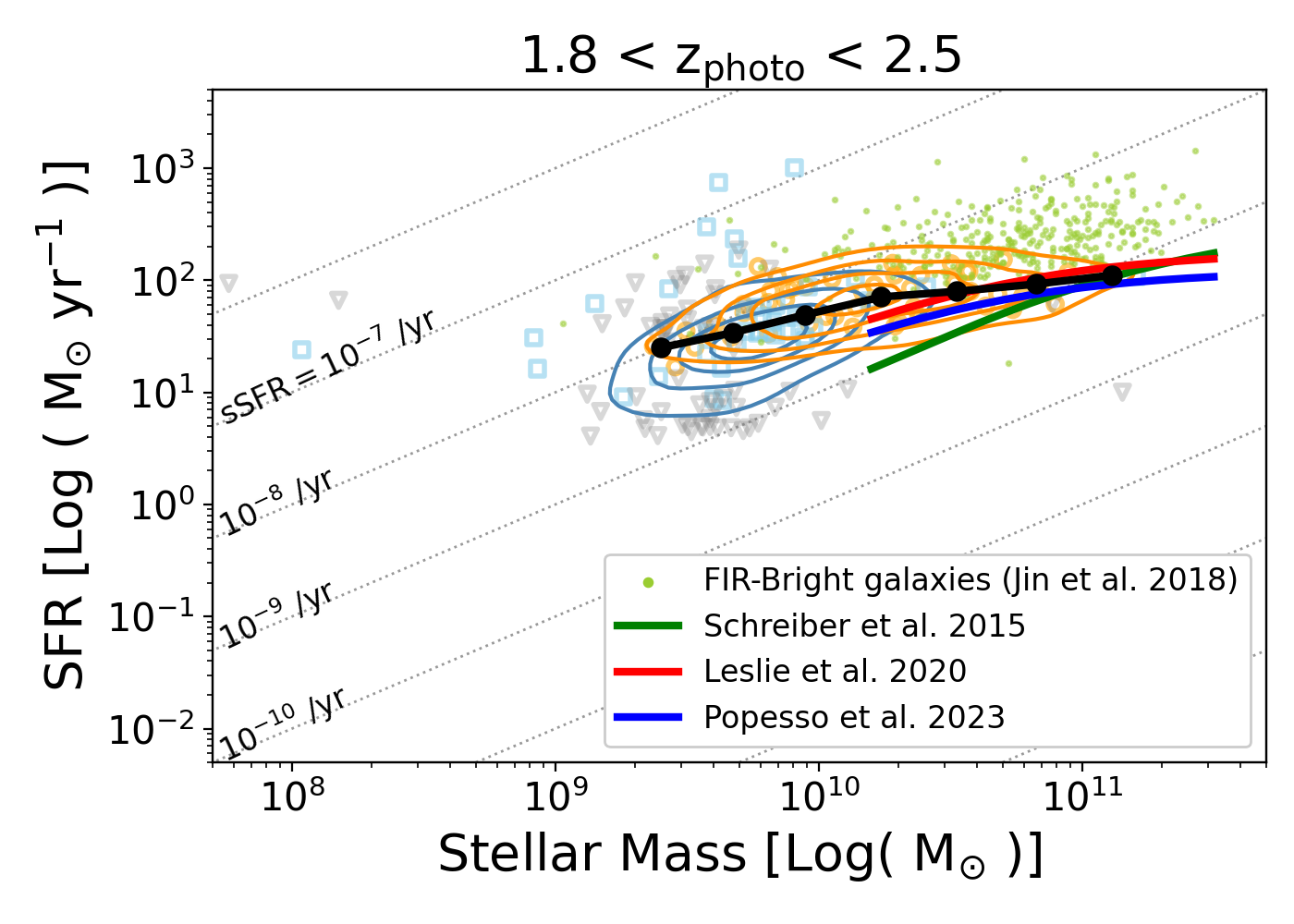}
\caption{  The Mass-SFR relation at the different redshift ranges.   The orange and blue contours are the distribution of the galaxies assigned to cells flagged as ``FIR clear" and ``FIR dim", respectively. The outermost contours contain 99$\%$ of the data.  The SFR of the galaxies are calculated using FIR luminosity inferred from our SOM.  The black lines are the median of the galaxy distribution accounting both the orange and blue contours.  The orange and blue squares are the median stellar mass and SFR labeled to the ``FIR clear" cells and ``FIR dim" cells, respectively. The grey triangles are the median stellar mass and SFR upper limits labeled to the ``No detection" cells.   We compare to the Mass-SFR relation  from \citet{Schreiber_2015, Leslie_2020, Popesso_2023}, and the individually detected FIR-bright galaxies (yellow-green points) in \citet{Jin_2018}.
\label{fig:M_SFR}}
\end{figure*}

We present the SFR-M relation with SFRs inferred from our SOM in different redshifts in Figure~\ref{fig:M_SFR}.  Here the SFR for individual galaxies is derived from the rescaled SFR$_{FIR}$ from the SOM, while the stellar mass are adopted from the COSMOS2020 CLASSIC catalog.  The orange contours are the distributions of the galaxies with SFR derived from the SOM cells flagged as ``FIR clear", and the blue contours are the distribution of the ``FIR dim" counterparts.  The cells (medians of each galaxy groups) flagged with ``FIR clear" and ``FIR dim" are also marked in orange and blue squares.  The median of the SFR binned along the stellar mass are marked as the black lines.  
We compare our result with the  SFR-M relation in the previous studies \citep{Schreiber_2015, Leslie_2020, Popesso_2023},  
and also show the individually detected galaxies in section~\ref{sec:photometry} (green points, plotted with the direct measured SFR from \citealt{Jin_2018}).  This individual detected galaxies indeed are the FIR bright outliers above the SFR-M relation.  The SFR-M relation is consistent with previous studies, and also shows the turnover behavior at the high-mass end. 

%We note that when we scale the star formation rate from the best-matching-cells, the scaling factors derived from the optical-to-NIR magnitude difference are roughly the same as the difference in stellar mass.  Therefore, galaxies assigned to the same best-matching-cell have similar sSFR in our derivation, regardless of their stellar masses.   The turnover of SFR-Mass relation, implying a change of sSFR, is caused by the change of the density and distribution of the best-matching-cells between the turnover mass. 

Previous studies binning galaxies by stellar mass and redshifts have more galaxies in each bin, therefore should obtain have deeper stacked images.  However, the mass-binning studies only go down to Log(M$_*$)$\simeq$9.5, since the galaxy population is incomplete at the low mass end.  Binning galaxies by color-SED does not avoid the incompleteness issue. However, galaxies assigned to the same best-matching cells are expected to have similar physical properties, thus the stacking result is less impacted by the incompleteness at the low mass end. Our SOM therefore provides a precise prediction according to different types of SEDs, and we are able to make prediction for galaxies with specific types of SED at the low mass end.

%%%%%%%%%%%%%%%%%%%%%%%%%%%%%%%%%%%%%%%%%%%%%%%%%%%%%%%%%%%%%%%%%%
%%%%%%%%%%%%%%%%%%%%%%%%%%%%%%%%%%%%%%%%%%%%%%%%%%%%%%%%%%%%%%%%%%
\section{Summary} \label{sec:summary}
%%%%%%%%%%%%%%%%%%%%%%%%%%%%%%%%%%%%%%%%%%%%%%%%%%%%%%%%%%%%%%%%%%
%%%%%%%%%%%%%%%%%%%%%%%%%%%%%%%%%%%%%%%%%%%%%%%%%%%%%%%%%%%%%%%%%%

We present the galaxy color-SFR$_{FIR}$ relation at $0<z<2.5$ in this work.  We first cluster 230,638 selected galaxies selected from the COSMOS2020 catalog onto a 40$\times$40 cell self-organizing map (SOM), using the adjacent colors $u-g$, $g-r$, $r-i$, $i-z$, $z-Y$, $Y-J$, $J-H$.   We then stack the Spitzer MIPS 24, 70 $\mu$m and Herschel PACS/SPIRE 100, 160, 250, 350 and 500$\mu$m images of galaxies that are assigned to the same best-matching cells on the SOM.  Stacking based on Optical-to-NIR colors will ensure that our FIR stacks are built from galaxies with similar intrinsic physical properties as opposed to stacking purely in stellar mass and redshift bins.  The summary of our results is as follows: 

\begin{enumerate}

    \item  The SOM trained on adjacent color SED successfully cluster the photometric redshift $z_{photo}$, with the redshift dispersion $<0.15$ for most of the cells, as well as  the physical scale, with the stellar mass dispersion $<0.4$ dex.  
    
    \item Features labeled to a SOM cell are expected to be similar to that of the neighboring cells.  We find that not only the features derived from the Optical-to-NIR SED (such as $z_{photo}$ and stellar mass) of a cell are similar to the surroundings, but the FIR signals also show similar structures.  This suggests that FIR SED can be extrapolated from the Optical-to-NIR SED. 
    
    \item We are able to measure the FIR luminosities of half of the cells on the SOM, and calibrated the FIR star formation rate.  We coordinate the stacked image cutouts to create a SOM mosaic, and plot the mosaic over the SOM labeled with the median of assigned galaxies' properties.   The distribution of cells with FIR detection on the SOM is most similar to that of the stellar mass and SFR labeled map, but not the sSFR.  

    \item We predict the SFR$_{\rm FIR}$ of individual galaxies by scaling the FIR luminosity of the stacked images with the Optical-to-NIR magnitude difference between the individual galaxies and the median value assigned to their best-matching-cell.  The SFR$_{\rm FIR}$ agrees with the SFR adopted from the COSMOS2020 catalog, with moderate scatter $\sim0.5$ dex. 

    \item Stacking images of mass-binning galaxies include various types of SED, and face an issue of incompleteness at the low mass end.  SED-binning ensure that galaxies participate in the stacking process have similar properties, therefore is less impacted by the incompleteness, provides a precise prediction according to different types of SEDs, and is also able to make predictions for galaxies with specific types of SED at the low mass end.  

\end{enumerate}

The summarized table of labeled features and measurements of the SOM, as well as the table of derived properties for individual galaxies can be downloaded here ({\it link will be added upon publication of this paper}).  The methodology of clustering galaxies by SED to enhance stacking results has great potential 
on analyzing the large-scale data that is or will be available from LSST, Euclid, SPHEREx, and Roman.

\section{Acknowledgment}

All the {\it COSMOS} imaging data used in this paper can be found in IRSA \citet{cosmos_hst}, and the COSMOS2020 catalog can be found in \citet{cosmos2020}.  
This work was supported by NASA’s Astrophysics Data Analysis Program (ADAP) under grant NNH21ZDA001N-ADAP. This research has made use of archival data provided by the NASA/IPAC Infrared Science Archive (IRSA), operated by JPL/Caltech, under contract with NASA.
The summarized table of our SOM and the table of SOM-derived properties for COSMOS galaxies are available on Zenodo: ({\it link will be added upon publication of this paper}).

\clearpage

\appendix

\section{The Bias in Different Stacking Methods }\label{sec:bias_in_simulation}

\subsection{The Mean and Median Stacking Methods}

Mean and median are widely used in the process of stacking. Although median is less sensitive to the present of outliers, the median stacking may create nontrivial bias depending on the underlying distribution and noise level of the images. Previous studies found that median stacking binned by redshift and stellar mass have a nontrivial bias to higher fluxes \citep{White_2007, Schreiber_2015, Leslie_2020}.  Since this work clusters the galaxies by the SED into much detailed groups but with lower numbers of images on stacking, we revisit the difference of the stacking process between mean and median.

We present a set of simulation results from Section~\ref{sec:simulation}, comparing the mean- and median-stacking methods using normalized images with the FIR intrinsic scatter being 0.1 dex and 0.3 dex, as well as the results without the normalization in the stacking process.  In every cell, we measure the flux densities $f_{\nu}$ of stacked images from galaxy images with noisy background, and compare that to the ``expected $f_{\nu}$'' which is measured from noiseless galaxy stacked images using the same stacking methods (mean to mean, median to median). In Figure~\ref{fig:model_mean_median} left panel, each points shows the median of the measured $f_{\nu}$ to expected $f_{\nu}$ ratios of stacked mock galaxies images, where the mock galaxies are sampled using the fluxes and populations of 30 randomly selected cells that are flagged as ``FIR clear" in our SOM.

We find that when the intrinsic scatter is small in the process with normalization, both median and means perform well. When the intrinsic scatter is large, the normalization does not concentrate the underlying FIR distribution, and performs similarly to the stacking method without normalization. In these cases, the mean stacking method successfully captures the expected fluxes of the mean-stacked images of the noiseless data, while the median stacking method is biased to higher fluxes compared to the median-stacked images of the noiseless data.

The mean-stacking method being able to capture the expected flux of the noiseless mean-stacked image, is unsurprising.  According to the central limit theorem, the distribution of the measured $f_{\nu}$ from the mean-stacked in the cells should be a Gaussian distribution centered on the expected $f_{\nu}$.   
The reason why the median-stacking method biases to higher values is that the underlying distributions of the FIR flux densities (modeled by the optical-to-NIR distribution, or the added random intrinsic scatter) are roughly described with a log-normal distribution. 
The log-normal distribution is right-skewed. 
The median is smaller than the mean in a right-skewed distribution.   

When we add the high-level symmetric noise centered at zero, it makes the distribution more symmetrical without changing the mean. Therefore, the median value of the noisy distribution moves closer to the mean and also increases.  
The flux density distribution of galaxies binned by redshifts and stellar masses is often described with the Schechter equation \citep{Schechter_1976}, which are also right-skewed. Therefore, the results of other binning using the median-stacking method will present similar bias.

\subsection{The Bias of the SOM cells}

\begin{figure}
\includegraphics[width=0.49\linewidth]{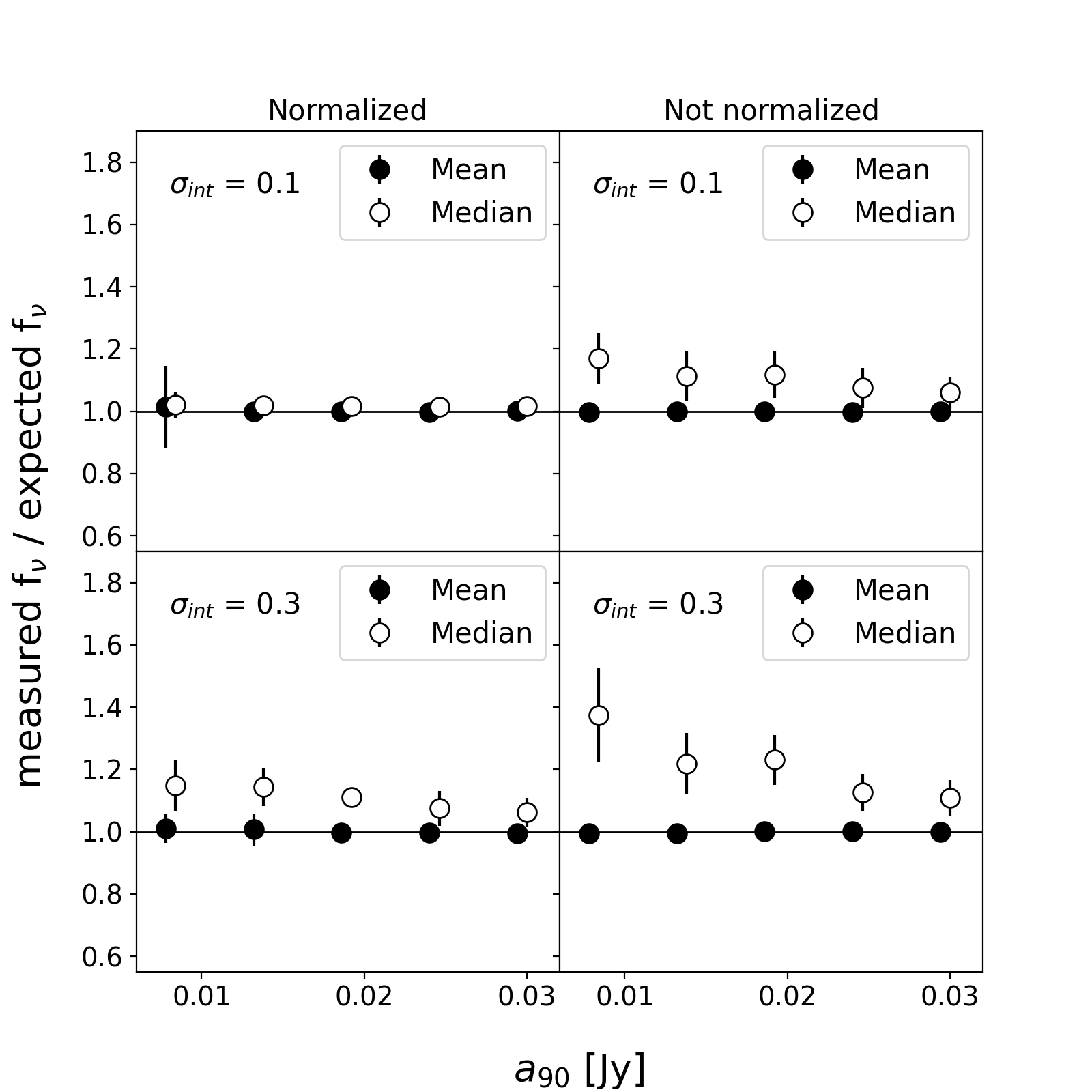}
\includegraphics[width=0.49\linewidth]{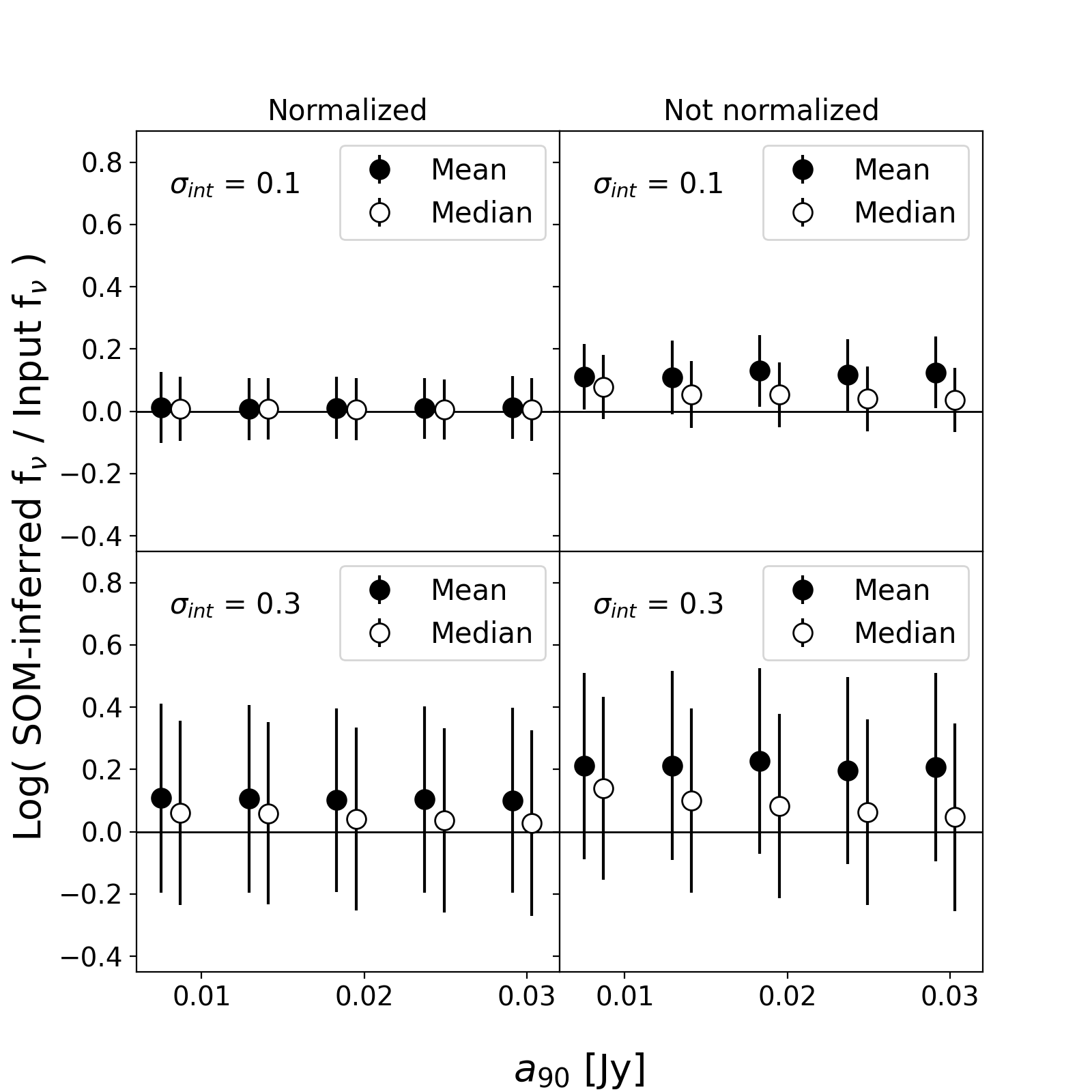}
\caption{  (Left):   The measured f$_\nu$ of stacked noisy images v.s. the expected $f_{\nu}$ of stacked noiseless images from our simulation.  The left and right columns shows the results with and without the normalization during the stacking, respectively. The points are the median $f_{\nu}$ of stack images generate from 30 randomly selected ``FIR clear" cells.  The black and white points stand for the mean- and median-stacking method, respectively.  The error bars are the 16th and 84th percentiles of the distributions.  
(Right):  Same as the left panel, but the logarithmic ratio of SOM-inferred flux density vs input flux densities of all the mock galaxies in the simulation.  
\label{fig:model_mean_median}} 
\end{figure}

The purpose of stacking the FIR images in this work is to boost the SNR and provide insight for the galaxy FIR photometry in each SOM cell. Therefore, we should choose between the mean- and median-stacking methods, of which the result best represents the galaxies in the SOM cells .  Even if the mean-stacking method can accurately reproduce the expected mean flux density, the mean value is still affected by the present of outliers and the uneven distribution . 
In Figure~\ref{fig:model_mean_median} right panel, we compare the individual input flux densities $f_{\nu}(gal)$ to the SOM prediction using the stacked images.  We find that median-prediction always has smaller bias than the mean-prediction, and when the normalization is involved, the bias of mean and median prediction are smaller, but the dispersion is roughly the same. 
Overall, the median-stacking on normalized images provides the most robust prediction. 

The uncertainties of the photometry labeled to each SOM cell are also shown in Figure~\ref{fig:model_mean_median} left panel.  If $\sigma_{int}$ is small, the photometry of the cell is unbiased.  If $\sigma_{int}$ is large (comparable to the optical-to-NIR scatter $\sim$0.3 dex), the photometry of the cell is slightly overestimated, and the bias increases for the cells representing fainter galaxies.  The uncertainty of the SOM prediction is approximately $\sigma = \sqrt{\sigma_{int}^2 + \sigma_{obs}^2}$, where $\sigma_{obs}$ is the uncertainty in observation.

\bibliography{sample631}{}
\bibliographystyle{aasjournal}

\clearpage



\end{document}